%% file: paper.tex
\documentclass[sigconf]{acmart}

\PassOptionsToPackage{protrusion, expansion}{microtype} 

\input{tex/00_preamble.tex}

\usepackage{array}

\copyrightyear{2022} 
\acmYear{2022} 
\setcopyright{rightsretained} 

\acmConference[HPDC '22]{Proceedings of the 31st International Symposium on High-Performance Parallel and Distributed Computing}{June 27-July 1, 2022}{Minneapolis, MN, USA}
\acmBooktitle{Proceedings of the 31st Int'l Symposium on High-Performance Parallel and Distributed Computing (HPDC '22), June 27-July 1, 2022, Minneapolis, MN, USA}
\acmDOI{10.1145/3502181.3531458}
\acmISBN{978-1-4503-9199-3/22/06}

\usepackage{etoolbox}
\makeatletter
\patchcmd{\maketitle}{\@copyrightpermission}{
   \begin{minipage}{0.3\columnwidth}
     \href{http://creativecommons.org/licenses/by/4.0/}{\includegraphics[width=0.90\textwidth]{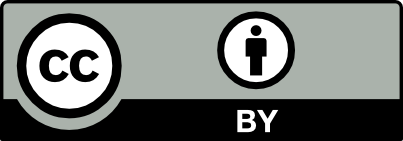}}
   \end{minipage}\hfill
   \begin{minipage}{0.7\columnwidth}
     \href{http://creativecommons.org/licenses/by/4.0/}{This work is licensed under a Creative Commons Attribution International 4.0 License.}
   \end{minipage}
  
   \vspace{5pt}
}{}{}

\makeatother

\begin{document}
\fancyhead{}

\title{TAC: Optimizing Error-Bounded Lossy Compression for Three-Dimensional Adaptive Mesh Refinement Simulations}

\input{ACM_author}

\input{abstract}



\maketitle


\newcommand{\kdtree}{$k$-d tree}

\setlength{\textfloatsep}{6pt}

\input{tex/01_introduction}
\input{tex/02_background}
\input{tex/03_design}

\input{tex/04_evaluation}

\input{tex/06_conclusion}
\input{tex/99_acknowledge}

\newpage
\bibliographystyle{ACM-Reference-Format}
\bibliography{refs.bib}

\end{document}

%% file: tex/00_preamble.tex
\usepackage{graphicx}

\usepackage{amsmath, amsthm}
    
\usepackage{graphicx, xcolor} 
\usepackage{hyperref, url}
\usepackage{adjustbox}
\usepackage{booktabs, multirow}
\usepackage{caption}
\usepackage{subcaption}
\usepackage{enumitem}
\usepackage{lipsum}
\usepackage{setspace}

\usepackage[ruled, linesnumbered]{algorithm2e}
\SetKwComment{Comment}{/* }{ */}
\usepackage{algpseudocode}
\algnewcommand{\algorithmicand}{\textbf{ and }}
\algnewcommand{\AND}{\algorithmicand}
\algnewcommand{\algorithmicor}{\textbf{ or }}
\algnewcommand{\OR}{\algorithmicor}

\SetCommentSty{mycommfont}

\usepackage{footnote}
\makesavenoteenv{tabular}
\makesavenoteenv{table}

\usepackage[zerostyle=d]{newtxtt}



%% file: ACM_author.tex
\settopmatter{authorsperrow=4}

\newcommand{\AFFIL}[4]{%
    \affiliation{%
        \institution{\small #1}
        \city{#2}\state{#3}\country{#4}
    }
    }
    
\author{Daoce Wang}{\AFFIL{Washington State University}{Pullman}{WA}{USA}}
\email{daoce.wang@wsu.edu}

\author{Jesus Pulido}{\AFFIL{Los Alamos National Laboratory}{Los Alamos}{NM}{USA}}
\email{pulido@lanl.gov}

\author{Pascal Grosset}{\AFFIL{Los Alamos National Laboratory}{Los Alamos}{NM}{USA}}
\email{pascalgrosset@lanl.gov}

\author{Sian Jin}{\AFFIL{Washington State University}{Pullman}{WA}{USA}}
\email{sian.jin@wsu.edu}

\author{Jiannan Tian}{\AFFIL{Washington State University}{Pullman}{WA}{USA}}
\email{jiannan.tian@wsu.edu}

\author{James Ahrens}{\AFFIL{Los Alamos National Laboratory}{Los Alamos}{NM}{USA}}
\email{ahrens@lanl.gov}

\author{Dingwen Tao}{\AFFIL{Washington State University}{Pullman}{WA}{USA}}
\authornote{Corresponding author: Dingwen Tao, School of Electrical Engineering and Computer Science, Washington State University, Pullman, WA 99163, USA.}
\email{dingwen.tao@wsu.edu}

%% file: abstract.tex
\begin{abstract}
Today's scientific simulations require a significant reduction of data volume because of extremely large amounts of data they produce and the limited I/O bandwidth and storage space. Error-bounded lossy compression has been considered one of the most effective solutions to the above problem. However, little work has been done to improve error-bounded lossy compression for Adaptive Mesh Refinement (AMR) simulation data. Unlike the previous work that only leverages 1D compression, in this work, we propose to leverage high-dimensional (e.g., 3D) compression for each refinement level of AMR data. To remove the data redundancy across different levels, we propose three pre-process strategies and adaptively use them based on the data characteristics. Experiments on seven AMR datasets from a real-world large-scale AMR simulation demonstrate that our proposed approach can improve the compression ratio by up to 3.3$\times$ under the same data distortion, compared to the state-of-the-art method. In addition, we leverage the flexibility of our approach to tune the error bound for each level, which achieves much lower data distortion on two application-specific metrics. 
\end{abstract}

%% file: tex/01_introduction.tex
\section{Introduction}
\label{sec:introduction}

\paragraph{Motivation} The increase in supercomputer performance over the last few years has been insufficient to solve many challenging modeling and simulation problems. For example, the complexity of solving evolutionary partial differential equations (PDEs) scales as $\Omega(n^4)$, where $n$ is the number of mesh points per dimension. Thus, the performance improvement of about three orders of magnitudes over the past 30 years has meant just a 5.6$\times$ gain in spatio-temporal resolution~\cite{burstedde2008towards}. To address this issue, many high-performance computing (HPC) simulation packages~\cite{dubey2014survey} (such as AMReX~\cite{zhang2019amrex} and Athena++~\cite{stone2020athena++}) use Adaptive Mesh Refinement (AMR)---which applies computation to selective regions of most interest---to increase resolution. Compared to the method where a high resolution is applied everywhere, the AMR method can greatly reduce the computational complexity and storage overhead; thus, it is one of most widely used frameworks for many HPC applications~\cite{almgren2013nyx, runnels2021massively, whitman2018simulation, sverdrup2018highly} in various science and engineering domains. 

Although AMR can save storage space to some extent, AMR applications running on supercomputers still generate large amounts of data, making the data transmission and storage challenging. 
For example, one Nyx simulation~\cite{nyx} with a resolution of $4096^3$ (i.e., $0.5 \times 2048^3$ mesh points in the coarse level and $0.5 \times 4096^3$ in the fine level ) can generate up to 1.8 TB of data for a single snapshot; a total of 1.8 PB of disk storage is needed assuming running the simulation 5 times with 200 snapshots dumped per simulation.
Therefore, reducing data size is necessary to lower the storage overhead and I/O cost and improve the overall application performance for large-scale AMR applications running on supercomputers. 

A straightforward way to address this issue is to use data compression. However, traditional lossless compression techniques such as GZIP~\cite{gzip} and Zstandard~\cite{zstd} can only provide a compression ratio up to 2 for scientific data~\cite{son2014data}. On the other hand, a new generation of lossy compressors which can provide a strict error control (called ``error-bounded'' lossy compression) has been developed, such as SZ~\cite{sz16,sz17,sz18}, ZFP~\cite{zfp}, MGARD~\cite{ainsworth2017mgard}, and TTHRESH~\cite{ballester2019tthresh}. Using those error-bounded lossy compressors, scientists can achieve relatively high compression ratios while minimizing the quality loss of reconstructed data and post analysis, as demonstrated in many prior studies~\cite{cappello2019use, jin2020understanding, grosset2020foresight, lu2018understanding, baker2014methodology, baker2017toward, gok2018pastri, wu2019full}. 

\textit{Limitation of state-of-the-art approach.} Only a few existing contributions have investigated error-bounded lossy compression for AMR applications and datasets. A common approach is to generate uniform resolution data by up-sampling the coarse-level data and merging them with the finest-level data, and then to perform compression on the merged data. However, this approach introduces redundant information to the data, which will significantly degrade the compression ratio, especially when the up-sampling rate is high or there are multiple coarse levels to up-sample. Recently, 
Luo~\textit{et~al.} introduced zMesh~\cite{zMesh}, a technique that groups data points that are mapped to the same or adjacent geometric coordinates such that the dataset is smoother and more compressible. However, since zMesh maps data points from different AMR levels to adjacent geometric coordinates and generates a 1D array, it cannot adopt 3D compression which most HPC simulations use. 
Moreover, zMesh is designed only for \textcolor{black}{patch-based} AMR applications. The \textcolor{black}{patch-based} AMR structure saves the data blocks that will be refined at the next level in the current level redundantly. While the state-of-the-art AMR framework AMReX provides quadtree/octree-based structure besides \textcolor{black}{patch-based} structure~\cite{amrstructure}, many newly developed AMR applications such as Nyx adopt the tree-based structure to avoid redundancy by only saving each data point in the level of its finest refinement. For this scenario, the reorganization approach proposed by zMesh may not improve the data smoothness appropriately (will be demonstrated in Section~\ref{sec:evaluation}).

\textit{Key contributions.} To solve these issues, we propose an approach (called \textsc{TAC}) to optimize error-bounded \underline{T}hree-dimensional \underline{A}MR lossy \underline{C}ompression.
Specifically, we propose to adopt 3D compression for each AMR level. However, each level may contain many empty regions (i.e., zero blocks), where data points are saved in other levels; these empty regions (zero blocks) significantly decrease the data smoothness/compressibility and increase the data size (hence reduce the compression ratio). 
Thus, we propose to either remove these empty regions or partially pad them with appropriate values, based on the density of empty regions.
Furthermore, we 
propose an optimization to reduce the time cost of removing empty regions. 
Finally, we evaluate \textsc{TAC} on seven datasets and compare it with the state-of-the-art approach.
Our main contributions are summarized as follows. 





\begin{itemize}[topsep=0pt,partopsep=1ex,parsep=0pt]
    \item We propose to leverage 3D compression to compress each level of an AMR dataset separately. We propose a hybrid compression approach based on the following three pre-process strategies and data characteristics (e.g., data density). 
    \item For sparse AMR data, we propose an optimized sparse tensor representation to efficiently remove empty regions.
    \item To reduce the time overhead of removing empty regions, we propose an optimization based on the enhanced \kdtree.
    \item For dense AMR data, we propose a padding approach to improve the smoothness and compressibility. 
    \item We tune the error bound for each AMR level for Nyx cosmology simulation, which improves the compression quality in terms of two application-specific post-analysis metrics.
    \item Experiments show that, compared to the state-of-the-art approach zMesh, \textsc{TAC} can improve the compression ratio by up to \textcolor{black}{3.3$\times$} under the same data distortion on the tested real-world datasets. 
\end{itemize}

\textit{Experimental methodology and artifact availability.}
We evaluate \textsc{TAC}
on seven datasets from two real-world AMR simulation runs. The AMR simulations are well-known, open-source cosmology simulations---Nyx~\cite{nyx}. We compare \textsc{TAC} with three baselines including zMesh using generic metrics such as compression ratio and peak signal-to-noise ratio (PSNR) and application-specific metrics such as power spectrum and halo finder. Our code and datasets are available at \url{https://github.com/hipdac-lab/3dAMRcomp}.

\textit{Limitations of the proposed approach.}
Compared with the approach that up-samples the coarse-level data and then compresses the data with uniform resolution (denoted by ``3D baseline''), \textsc{TAC} provides much better compression performance (i.e., rate-distortion), when the finest level of the AMR dataset has a relatively low density.
However, when the finest level has a relatively high density, \textsc{TAC} is slightly worse than the 3D baseline. We will discuss this limitation in detail in Section~\ref{subsec:evaRateDis}.

In Section~\ref{sec:background}, we present background information about error-bounded lossy compression, AMR method, \kdtree, and related work on AMR data compression. In Section~\ref{sec:design}, we describe our proposed pre-process strategies and hybrid compression. In Section~\ref{sec:evaluation}, we show the experimental results on different AMR datasets. In Section~\ref{sec:conclusion}, we conclude our work and discuss the future work.







%% file: tex/02_background.tex
\section{Background and Related Work} 
\label{sec:background}
\textcolor{black}{In this section, we introduce background information about lossy compression for scientific data, AMR method and data, classic \kdtree{} used in particle data compression, and discuss the state-of-the-art method of AMR data compression and remaining challenges.}

\subsection{Lossy Compression for Scientific Data} 
There are two main categories for data compression: lossless and lossy compression. Compared to lossless compression, lossy compression can offer much higher compression ratio by trading a little bit of accuracy. There are some well-developed lossy compressors for images and videos such as JPEG~\cite{wallace1992jpeg} and MPEG~\cite{le1991mpeg}, but they do not have a good performance on the scientific data because they are mainly designed for integers rather than floating points. 

In recent years there is a new generation of lossy compressors that are designed for scientific data, such as SZ~\cite{sz16, sz17, sz18}, ZFP~\cite{zfp}, MGARD~\cite{ainsworth2017mgard}, and TTHRESH~\cite{ballester2019tthresh}.
These lossy compressors provide parameters that allow users to finely control the information loss introduced by lossy compression. 
Unlike traditional lossy compressors such as JPEG~\cite{wallace1992jpeg} for images (in integers), SZ, ZFP, MGARD, and TTHRESH are designed to compress floating-point data and can provide a strict error-controlling scheme based on the user's requirements.
Generally, lossy compressors provide multiple compression modes, such as error-bounding mode and fixed-rate mode.
Error-bounding mode requires users to set an error type, such as the point-wise absolute error bound and point-wise relative error bound, and an error bound level (e.g., $10^{-3}$). The compressor ensures that the differences between the original data and the reconstructed data do not exceed the user-set error bound level.

In this work, we focus on the SZ lossy compression (2021 R\&D 100 Award Winner~\cite{rd100}) because SZ typically provides higher compression ratio than ZFP~\cite{lu2018understanding,zhao2021optimizing} and higher (de)compression speeds than MGARD~\cite{zhao2021optimizing,liang2021error} and TTHRESH~\cite{ballester2019tthresh}. 
SZ is a prediction-based error-bounded lossy compressor for scientific data. It has three main steps: (1) predict each data point's value based on its neighboring points by using an adaptive, best-fit prediction method; (2) quantize the difference between the real value and predicted value based on the user-set error bound; and (3) apply a customized Huffman coding and lossless compression to achieve a higher ratio.

\subsection{AMR Method and AMR data}

\begin{figure}[t]
     \centering
     \begin{subfigure}[t]{0.28\linewidth}
         \centering
         \includegraphics[width=\linewidth]{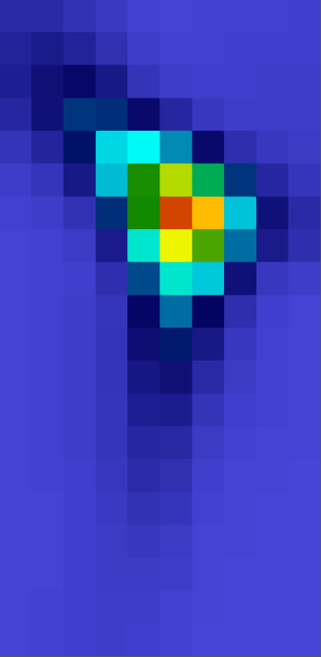} 
        \vspace{-5mm}
         \label{fig:vist2}
     \end{subfigure}
     \begin{subfigure}[t]{0.28\linewidth}
         \centering
         \includegraphics[width=\linewidth]{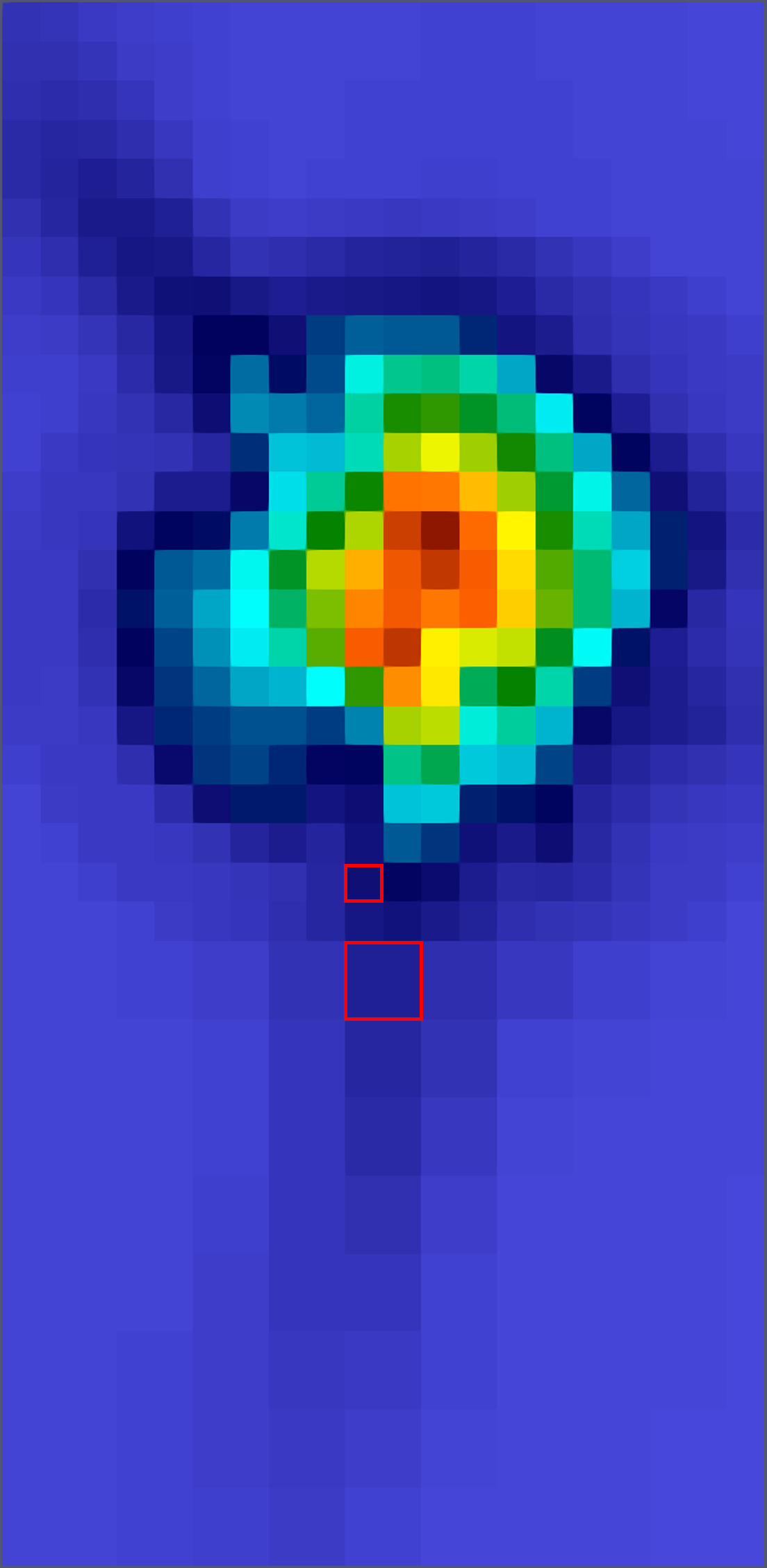}
        \vspace{-5mm}
         \label{fig:vist3}
     \end{subfigure}
      \begin{subfigure}[t]{0.28\linewidth}
         \centering
         \includegraphics[width=\linewidth]{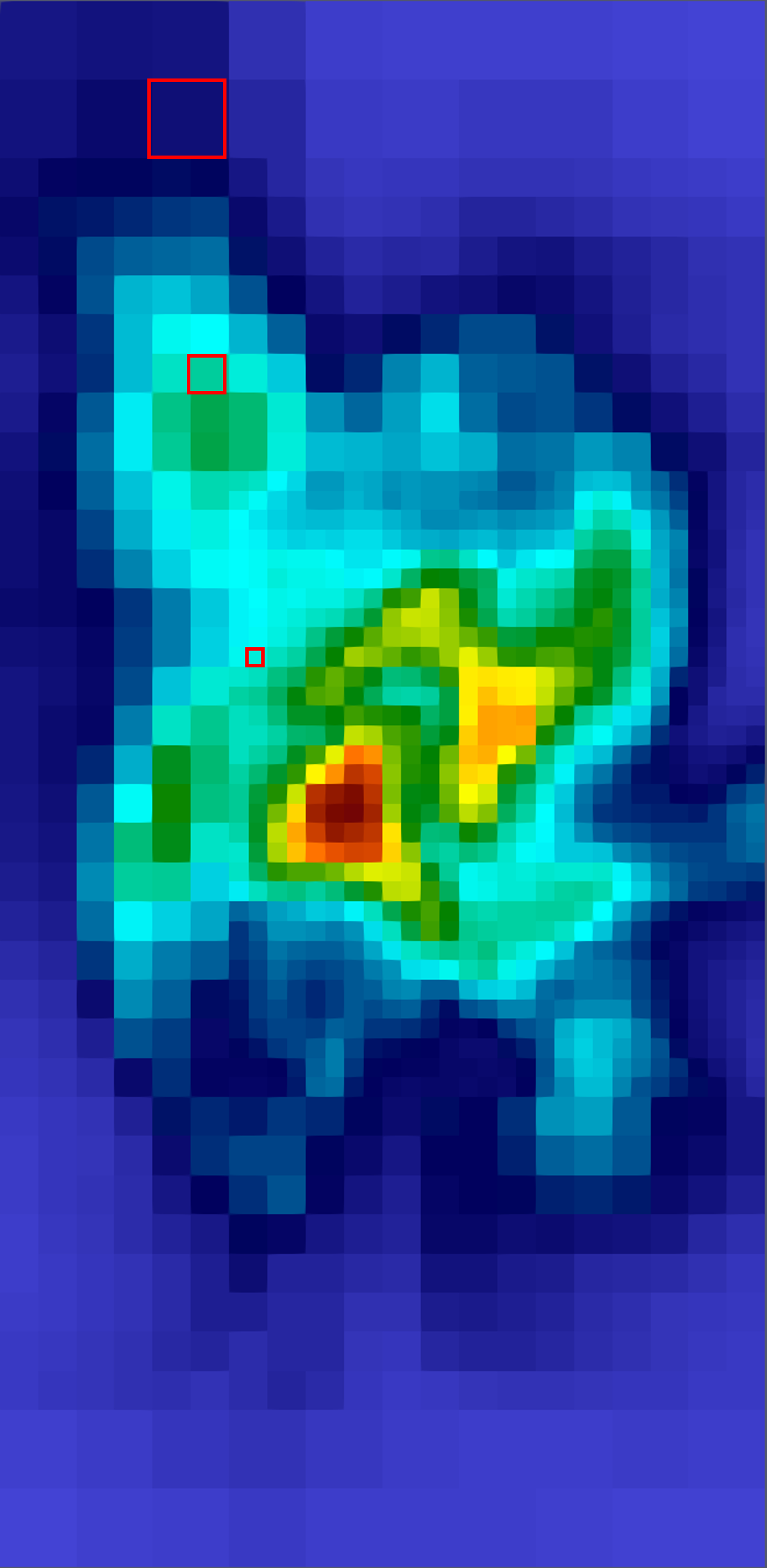}
        \vspace{-5mm}
         \label{fig:vist3}
     \end{subfigure}
        \vspace{-2mm}
        \caption[t]{Visualization (one zoom-in 2D slice) of three key timesteps generated from an AMR-based cosmology simulation. The grid structure changes with the universe's evolution. The red boxes indicate different resolutions within one AMR level.}
        \vspace{-2mm}
        \label{fig:visamr}
\end{figure}

\label{sec:amr}
AMR is a method of adapting the accuracy of a solution (e.g., solving hydrodynamics equations) by using a non-uniform grid to increase computational and storage savings while still achieving the desired accuracy. AMR applications change the mesh or spatial resolution based on the level of refinement needed by the simulation and use \textit{finer mesh in the regions with more importance/interest} and \textit{coarser mesh in the regions with less importance/interest}. \textcolor{black}{Figure \ref{fig:visamr} shows that during an AMR run, the mesh will be refined when 
the value meets the refinement criteria, e.g., refining a block when its norm of the gradients or maximum value is larger than a threshold \cite{lanlamr}.}

\begin{figure}[h]
    \centering 
    \vspace{-2mm}
    \includegraphics[width=0.98\columnwidth]{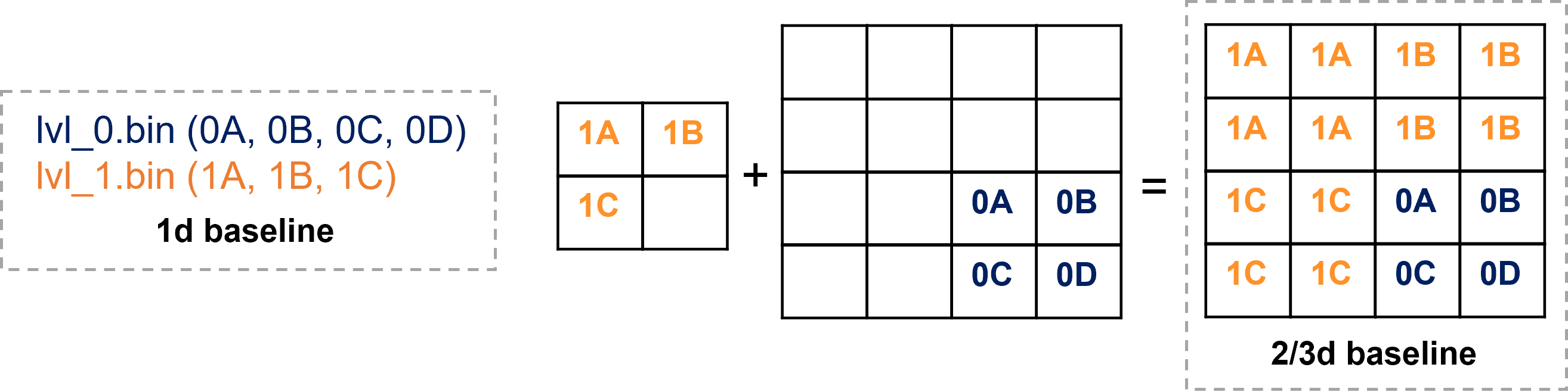}
    \vspace{-4mm}
    \caption{A typical example of AMR data storage and usage.}
    \vspace{-3mm}
    \label{fig:base_ex}
\end{figure} 

Clearly, the data generated by an AMR application are hierarchical data with different resolutions. The data of each AMR level are usually stored separately (e.g., in a 1D array). For example, Figure~\ref{fig:base_ex} (left) shows a simple example of two-level AMR data; ``0'' means high resolution (the fine level) and ``1'' for low resolution (the coarse level).
When the AMR data are needed for post analysis or visualization, users will typically covert the data from different levels to a uniform resolution. In the previous example, we will up-sample the data in the coarse level and combine it with the data in the fine level, as shown in Figure~\ref{fig:base_ex} (right).


\subsection{Existing AMR Data Compression}

\subsubsection{1D AMR Compression} 

The main challenge for AMR data compression is that the AMR data is comprehensive and hierarchical with different resolutions.
A naive approach is to compress the 1D data of each AMR level separately. However, this approach loses most of the topological/spatial information, which is critical for data compression.
zMesh~\cite{zMesh} is a state-of-the-art AMR data compression based on the 1D approach.
Different from the naive 1D approach, zMesh
re-organizes the 1D data based on each point's coordinate in the 2D layout; in other words, zMesh puts the points neighbored in the 2D layout closer in the 1D array.
It can increase the data smoothness/compressibility to benefit the following 1D compression such as SZ on the traditional \textcolor{black}{patch-based} ~\cite{cpu_rt} AMR data with redundancy. 
However, zMesh does not leverage high-dimensional compression, while many previous studies~\cite{sz17,zhao2020significantly} proved that leveraging more dimensional information (e.g., spatial/temporal information) can significantly improve the compression performance (e.g., compression ratio). Moreover, it only focuses on 2D \textcolor{black}{patch-based} AMR data. \textsc{TAC} aims to leverage high-dimensional data compression and supports 3D AMR data.

\vspace{-1mm}
\subsubsection{High-dimensional AMR Compression} 
Similar to the idea described in Section~\ref{sec:amr}, a straightforward way to leverage 3D compression on 3D AMR data is to compress different levels together by up-sampling coarse levels. 
However, this approach must
handle extra redundant data generated by the up-sampling process. 
As shown in Figure~\ref{fig:base_ex}, \textit{1A}, \textit{1B}, and \textit{1C} are redundant points in the compression. 
Note that the storage overhead of these redundant points will be higher when more data are in the coarse levels or up-sampling rate is higher, especially for 3D AMR data. 
This is because we only need to duplicate one point from the coarse level for 4 times for 2D AMR data but 8 times for 3D AMR data, with an up-sampling rate of 2. 
Another limitation of this approach is that it cannot apply different compression configurations (e.g., error bound) to different AMR levels, because after up-sampling all data points will have the same importance.
However, the purpose of using the AMR method is to set different interests to different AMR levels, so the error bound for each AMR level can be chosen adaptively based on the analysis. 

\subsection{$k$-D Tree for Particle Data Compression} 

\kdtree{}~\cite{kd1975} is a binary tree in which every node represents a certain space. Without loss of generality, for the 3D case, every non-leaf node in a \kdtree{} splits the space into two parts by a 2D plane associated with one of the three dimensions. 
The left subspace is associated with the left child of the node, while the right subspace is associated with the right child.
\kdtree{} is commonly used in particle data compression~\cite{Hoangldav21, kd-1, kd-2} to locate each particle and remove empty regions. Specifically, a \kdtree{} keeps dividing the space in between along one dimension until the space is empty or contains only one particle. 
We will propose to optimize the classic k-d tree and use it to remove empty regions and increase the data compressibility for each AMR level (to be detailed in Section~\ref{sec:kd}).


%% file: tex/03_design.tex
\section{Our Proposed Design}
\label{sec:design}

\begin{figure}[b]
    \centering 
    \vspace{-2mm}
    \includegraphics[width=0.99\columnwidth]{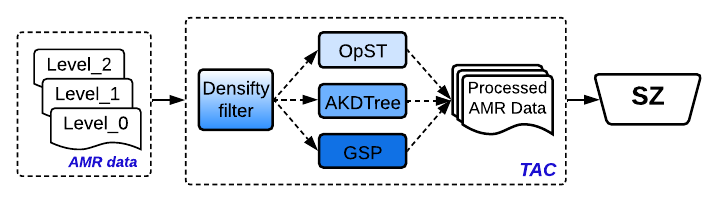}
     \vspace{-4mm}
    \caption{Workflow overview of our proposed \textsc{TAC}.}
    \label{fig:flow}
\end{figure} 

In this section, we propose a pre-process approach for AMR data to leverage high-dimensional data compression algorithms in each AMR level.
Specifically, we propose three pre-process strategies to mitigate the issue of irregular data distribution.
We also propose an adaptive approach to select the best-fit pre-process strategy based on the data characteristic (e.g., density) of each AMR level. 
\textcolor{black}{Figure~\ref{fig:flow} show the overview of our proposed \textsc{TAC}. It has a density filter that determines the best-fit pre-process strategy for each AMR level in the AMR dataset before compression. We will now illustrate our proposed three strategies in Section~\ref{sec:opst}, \ref{sec:kd}, and \ref{sec:gsp}, respectively.} 

\subsection{Optimized Sparse Tensor Representation for Low-density Data} \label{sec:opst}
\begin{figure}[t]
     \centering
     \begin{subfigure}[t]{0.49\linewidth}
         \centering
         \includegraphics[width=\linewidth]{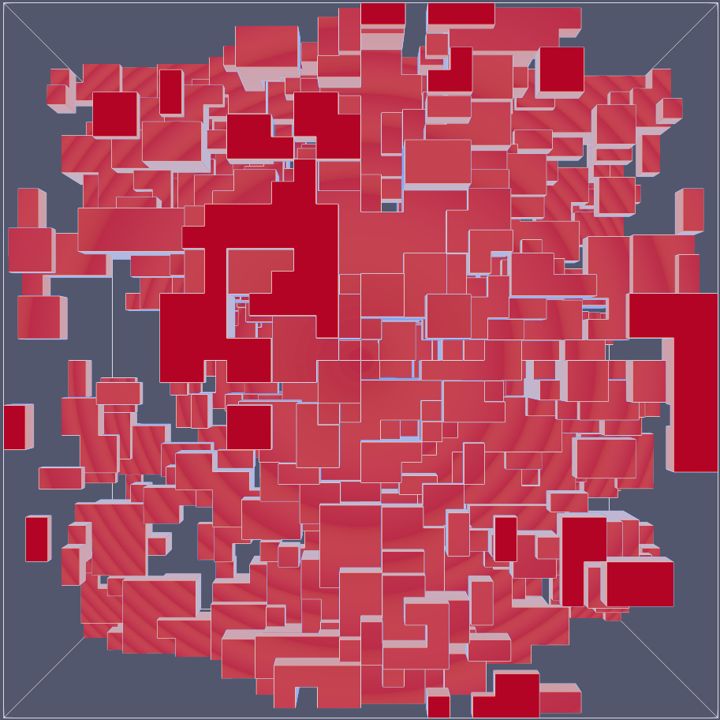}  
         \caption[t]{z10 fine level}
         \label{fig:dis_fine}
     \end{subfigure}
     \begin{subfigure}[t]{0.49\linewidth}
         \centering
         \includegraphics[width=\linewidth]{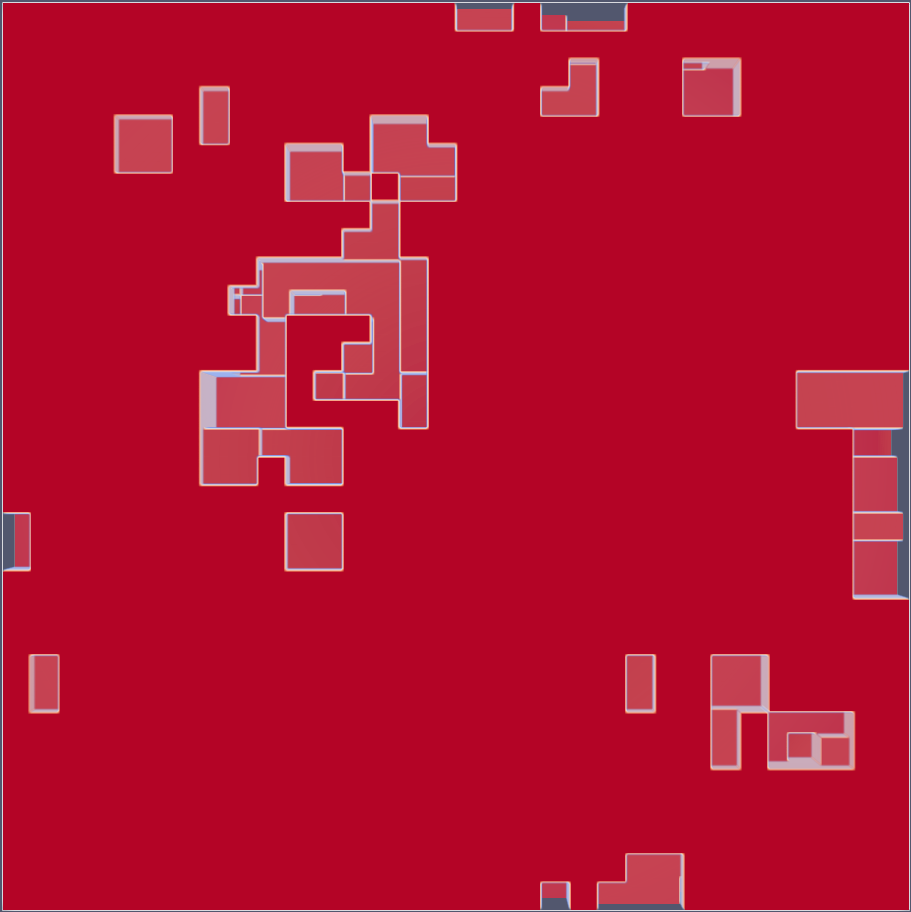}
         \caption{z10 coarse level}
         \label{fig:dis_coarse}
     \end{subfigure}
     \vspace{-2mm}
        \caption[t]{Visualization of data distributions of an example AMR data ``z10'', where z = redshift. Non-empty regions are shown in red.  
        }
        \label{fig:dis}
\end{figure}

To compress the AMR data in 3D, besides the aforementioned 3D baseline, we can also compress each level separately in 3d. 
However, in that way, the data will be split into multiple levels, 
and each level will have many empty regions and an irregular data distribution, as shown in Figure~\ref{fig:dis}. A naive solution to handle the irregular 3D data is to fill the empty regions with zeros and pass a large 3D block to the compressor. However, when most of the regions in the data are empty (e.g., about 77\% of the data is empty in Figure~\ref{fig:dis_fine}), we have to fill up many zeros, which would greatly increase the size of data for compression, resulting in a low compression ratio. 

\begin{figure}[h]
    \centering 
    \vspace{-2mm}
    \includegraphics[width=0.8\columnwidth]{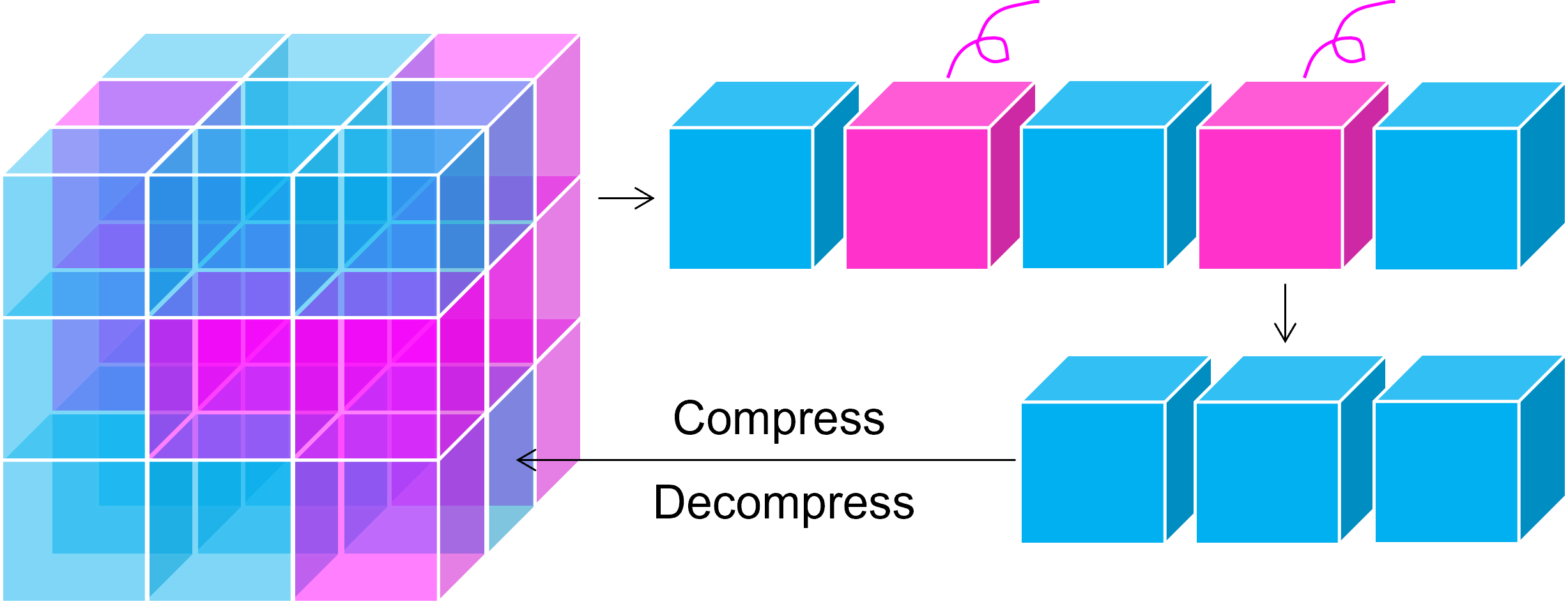}
     \vspace{-2mm}
    \caption{Workflow of the naive sparse tensor (NaST) method (empty regions marked in pink and non-empty regions marked in blue).}
    \vspace{-2mm}
    \label{fig:st-flow}
\end{figure}

To solve this issue, we propose to use a naive sparse-tensor-based approach (called \textbf{NaST}) to remove the empty regions, as shown in Figure~\ref{fig:st-flow}. NaST includes four main steps in the compression process: (1) partition the 3D data into multiple unit blocks, (2) remove the empty blocks, (3) linearize the remaining 3D blocks into a 4D array, and (4) pass the 4D array to the compressor. Note that in the decompression process, we will put the unit blocks from the decompressed 4D array back to the original data.


However, in order to completely remove the empty regions to form a sparse representation, the unit block size needs to be relatively small compared to the input data size (e.g., $16^3$ vs. $512^3$),
resulting in a high proportion of data on the boundary.
While boundary data have less information of neighboring data than non-boundary data, thus, it is harder for prediction-based compressors such as SZ to predict the boundary data values. 
As a result, the NaST method without optimizing the boundary data would have low compression performance.

\begin{figure}[t]
    \centering 
    \vspace{-2mm}
    \includegraphics[width=\columnwidth]{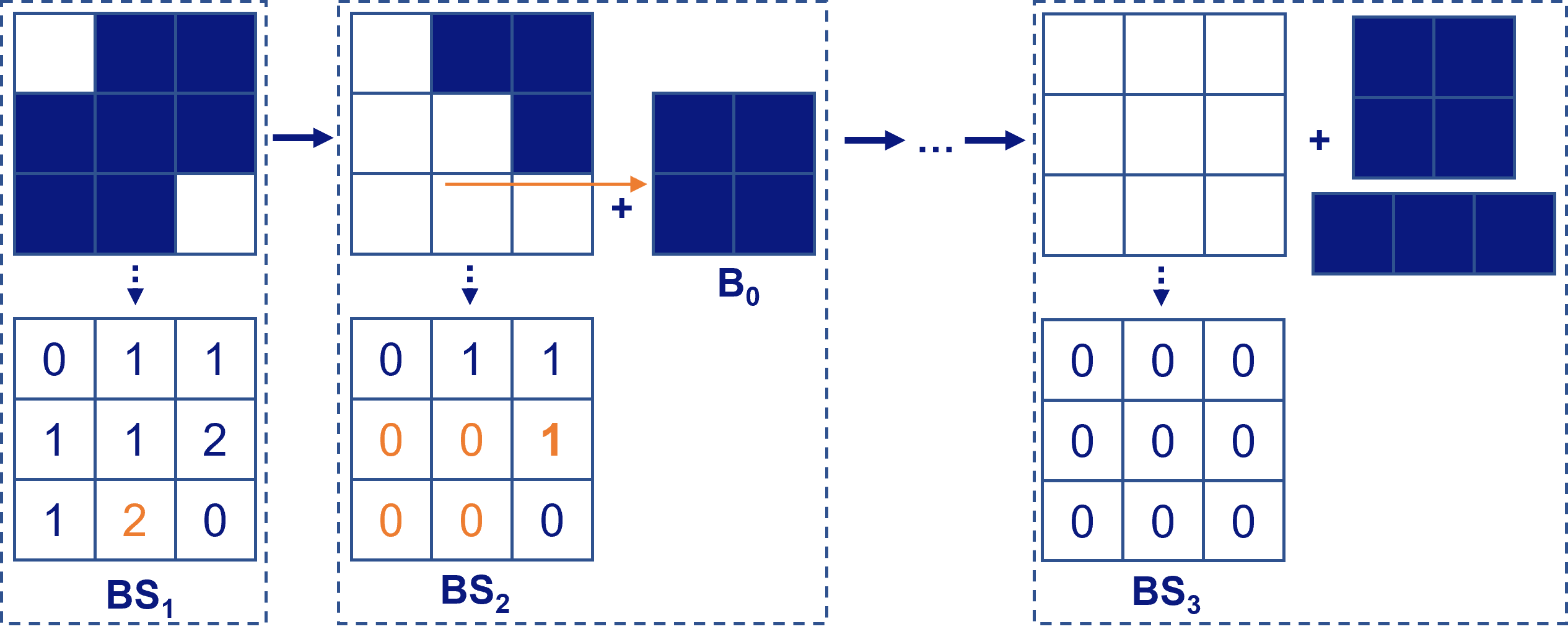}
     \vspace{-4mm}
    \caption{A 2D example of our proposed OpST approach. The sub-blocks are extracted according to our optimized sizes saved in $BS$. E.g., a 2-by-2 sub-block $B_0$ is extracted according to $BS_1[2][1]$.}
    \label{fig:opst-e}
\end{figure}

To address the above problem, we propose an optimized sparse tensor representation (called \textbf{OpST}) to effectively remove the empty regions as well as maintain a relatively large unit block size so as to reduce the portion of boundary data.
The detailed description of our algorithm can be found in Algorithm~\ref{alg:opst}.
We use a 2D example to demonstrate our approach, as illustrated in Figure~\ref{fig:opst-e}. 
Specifically, (1) we partition the data into many small unit blocks.
(2) For each unit block, we use the dynamic programming method to initiate an array $BS$ to save the dimension/size of the maximum square whose bottom-right corner is that unit block (line 6, will be discussed in the next paragraph). 
(3) We extract the sub-blocks (composing of multiple unit blocks) from the original data according to the sizes saved in $BS$ (lines 6 and 7). 
(4) Since the original data will be changed after the extraction, we need to partially update $BS$ based on \textit{maxSide} (will be discussed later).  
We loop (3) and (4) from the bottom-right corner to the top-left corner until the original data is empty.
(5) After extracting all the sub-blocks, we put them into multiple 3D arrays (to be compressed) based on their sizes. 
Note that the sub-blocks with the same size will be merged into the same array for easy compression.

When initializing the $BS$ in the step (2), we start with the $b'[i][j]$ with $i = 0$ or $j = 0$ (i.e., on the top-left edge), where $b'[\cdot][\cdot]$ are the unit blocks: if $b'[i][j]$ is empty, we will set $BS[i][j]$ to 0 otherwise 1. 
For the remaining unit blocks, if it is empty, $BS[i][j]$ will be 0; otherwise, $BS[i][j]$ will be set to 1 plus the minimum value among its three neighboring blocks (i.e., upper block, left block, and upper-left block).
In other words, we have $BS[i][j] = 1 + \min(BS[i][j-1], BS[i-1][j], BS[i-1][j-1])$ for the 2D case. For example, $BS_1[2][1]$ is 2 because all its upper-left neighbors are 1 (as shown in Figure~\ref{fig:opst-e}). However, both $BS_1[1][1]$ and $BS_2[1][2]$ can only reach 1 because one of their neighbors are set to 0, having no chance to form a sub-block with the size of 2. 

\begin{algorithm}[ht!]
\caption{Proposed Optimized Sparse Tensor Method}\label{alg:opst}
\KwIn {Sparse 3D data S}
\KwOut {multiple 4D array $D_n$}
\For{each unit block $b(x,y,z)$}{
  \If{$b(x,y,z)$ is non-empty}{
    \eIf{x is 0 \OR y is 0 \OR z is 0}{
        $BS(x,y,z) = 1$
      }{
        $BS(x,y,z) = \min( BS(x-1, y, z),~BS(x,y-1,z), ~BS(x,y,z-1), ~BS(x-1,y-1,z), ~BS(x,y-1,z-1), ~BS(x-1,y,z-1), ~BS(x-1,y-1,z-1)) + 1$
        \tcc*{BS(x,y,z) is the dimension size of the maximum cube whose bottom right rear corner is the unit block with index (x,y,z) in the original data}
        $maxSide = \max(maxSide, ~BS(x,y,z))$
      }
  }
}
\For{each unit block $b(x,y,z)$}{
    \If{$BS(x,y,z) \geq 1$}{
        $size = BS(x,y,z)$
        $D_{size} \leftarrow S((x-size:x) * blkSize,~ (y-size:y) * blkSize,~ (z-size:z)* blkSize)$
        \tcc*{put the sub-block to the according 4D array}
        $b(x-size:x,~y-size:y,~z-size:z) \leftarrow empty$
        $BS(x-size:x,~y-size:y,~z-size:z) = 0$
        $BS = updateBs(BS,~x,~y,~z,~ maxSide )$
     }
}
return $D_n$
\end{algorithm}

Moreover, as mentioned in the step (3), we need to update $BS$ after each extraction.
Specifically, for each sub-block we extract, we have to set its corresponding values in $BS$ to zeros. 
For instance, as shown in Figure~\ref{fig:opst-e}, after we extract a 2-by-2 sub-block $B_0$ at $BS_1[2][1]$, we need to set $BS_2[1][0]$, $BS_2[1][1]$, $BS_2[2][0]$, and $BS_2[2][1]$ to zeros.
In addition, we also need to recalculate a part of $BS$ (line 17 in Algorithm~\ref{alg:opst}) because the extraction could influence other $BS$ values. 
For example, we need to recalculate $BS_2[1][2]$ (marked in bold orange) after extracting $B_0$.
Note that this update is a partial update as the $BS$ values to be updated will be bounded by \textit{maxSide} which is the dimension size of the largest cube in the dataset (line 7). 

Similar to the NaST method, during decompression we will put the sub-blocks back to reconstruct the data based on the saved coordinates. Note that after our optimization, each sub-block size will be relatively large (e.g., $96^3$ versus the original data size of $512^3$), the metadata overhead of saving the coordinates of all the sub-blocks will be negligible (\textcolor{black}{e.g., 0.1\%}). 

\begin{figure}[h]
     \centering
     \begin{subfigure}[t]{0.49\linewidth}
         \centering
         \includegraphics[width=\linewidth]{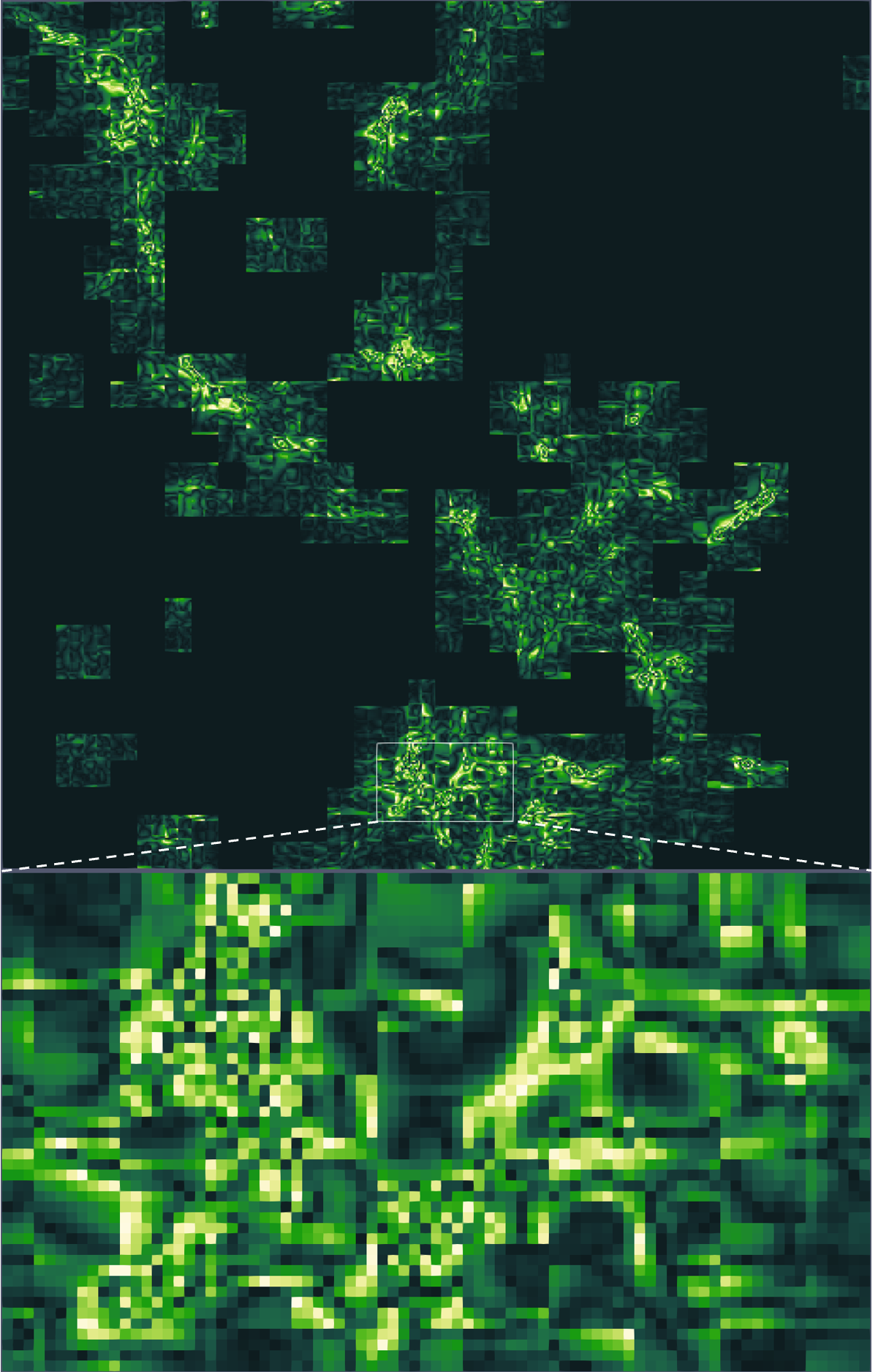}  
         \caption[t]{NaST (CR = 233.8, PSNR = 76.9 dB)}
         \label{fig:st-err}
     \end{subfigure}
     \vspace{-1mm}
     \begin{subfigure}[t]{0.49\linewidth}
         \centering
         \includegraphics[width=\linewidth]{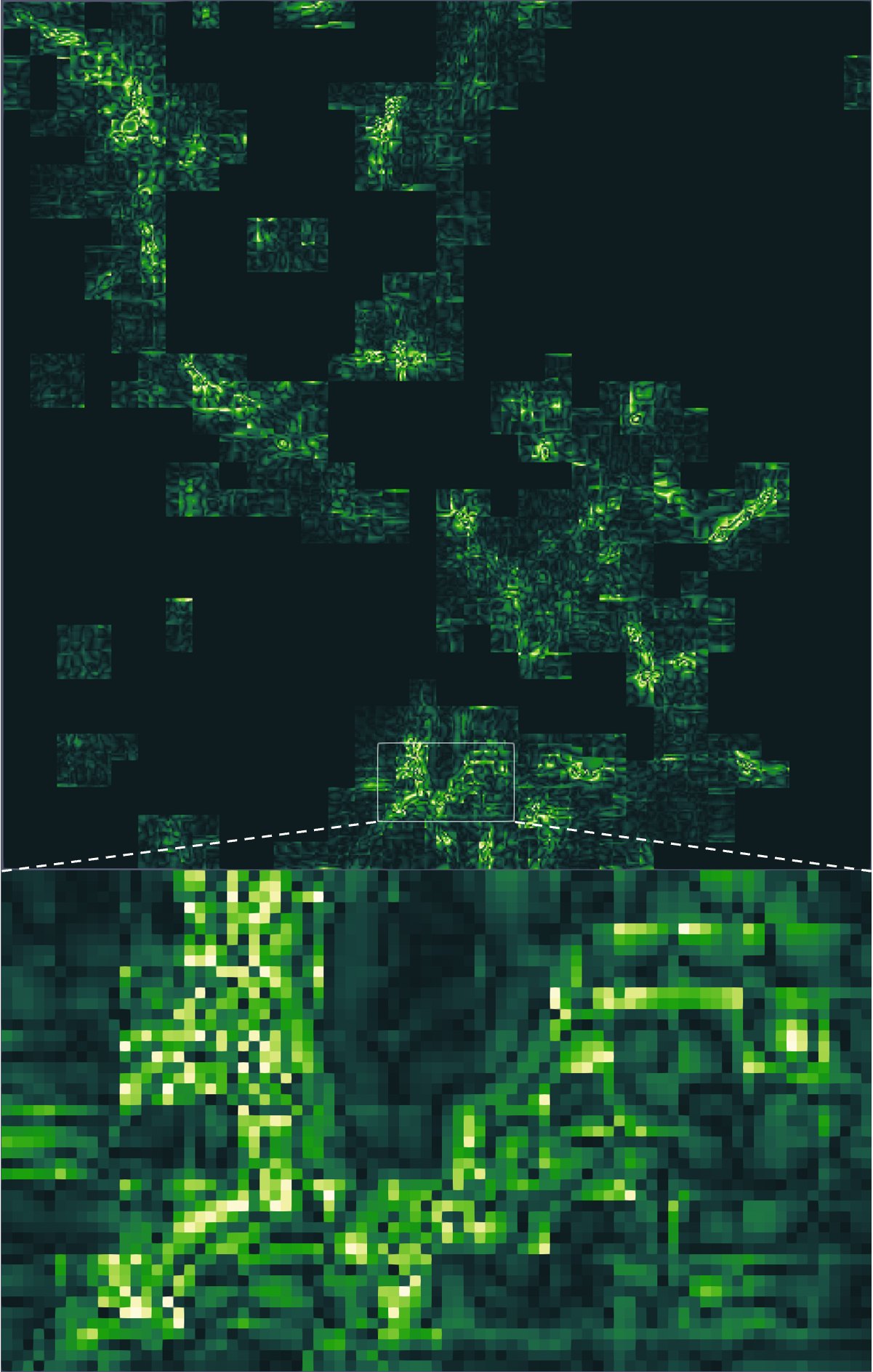}
         \caption{OpST (CR = 241.1, PSNR = 77.8 dB)}
         \label{fig:opst-err}
     \end{subfigure}
      \vspace{-1mm}
        \caption[t]{Visual comparison (one slice) of compression errors of two approaches using SZ based on Nyx ``baryon density'' field (i.e., z10's fine level, 23\% density). Brighter means higher compression error. The error bound is the relative error bound of $4.8\times10^{-4}$.}
        \label{fig:st_opst}
\end{figure}

Finally, we show a visual comparison of the compression quality between NaST and OpST in Figure~\ref{fig:st_opst}.
Note that both use the same compressor with the same error bound. Brighter means more error. 
We can observe that compared to the NaST method, OpST can significantly reduce the overall compression error, especially for the data points on the boundary.
It is worth noting that even with lower error, our OpST can still provide a higher compression ratio than NaST. 
This is because our proposed optimization will generate larger sub-blocks, which provide more information for prediction-based lossy compressors such as SZ to achieve better rate-distortion.
A detailed evaluation will be shown in Section~\ref{sec:evaluation}.



\subsection{Adaptive $k$-D Tree for Medium-density Data}
\label{sec:kd}

The OpST approach proposed for low-density data, however, has a high computation overhead, especially when the data is relatively dense. 
This is because, on one hand, OpST needs to update \textit{BS} based on \textit{maxSide} for each extraction of a sub-block, while the larger the \textit{maxSide}, the more values in \textit{BS} that need to be updated; on the other hand, \textit{maxSide} is the dimension size of the largest non-empty cube in the dataset, which is highly related to the density of the dataset.
Thus, the time complexity of OpST can be expressed as $O(N^2 \cdot d)$, where $N$ is the unit block number and $d$ is the density. 
Note that here density describes how dense the data is. 
For example, the density of 77\% means that 23\% of the data is empty.
Clearly, when the density of an AMR level is relatively high, using OpST for compression will be relatively time-consuming. 

\begin{figure}[h]
    \centering 
    \includegraphics[width=0.95\columnwidth]{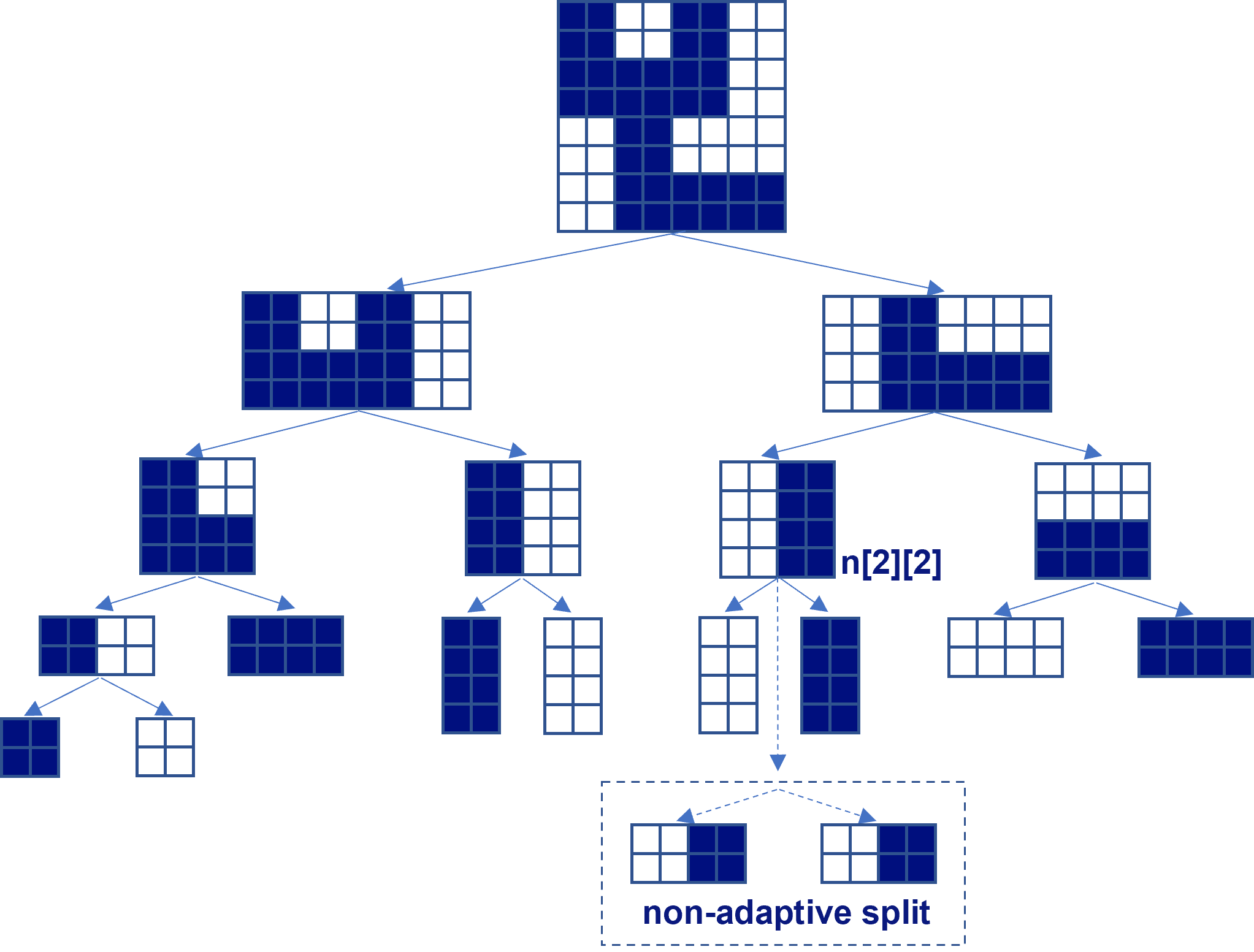}
    \caption{2d Example of adaptive k-d tree, the sub-block will be adaptively split to in order to effectively remove the empty region as well as get bigger full sub-block.}
    \label{fig:kde}
\end{figure} 

\begin{algorithm}[b]
\caption{Dynamic $k$-D Tree}\label{alg:kd}
\KwIn {data block $d$, counts information}
\KwOut {$k$-d tree}
node.count $\leftarrow$ counts information\;
\eIf{$d$ is empty \OR $d$ is full}{
    continue \tcc*{stop splitting}
}{
    \uIf{$d$ is a cube}{
    split d equally into 8 oct-blocks: $s_1,\cdots,s_8$\;
    get the counts $c_1,...c_8$ for $s_1,\cdots,s_8$\;
    find the maxDiff partition $d_1$,$d_2$\;
    node.left\ \ = AKDTree ($d_1$, four $c_i$ of $d_1$)\;
    node.right = AKDTree ($d_2$, four $c_i$ of $d_2$)\;
    }
    \uElseIf{$d$ is a flat cuboid}{
      get the counts $c_1,\cdots,c_4$ from counts information\;
      find the maxDiff partition $d_1$, $d_2$\;
      node.left\ \ = AKDTree ($d_1$, two $c_i$ of $d_1$)\;
      node.right = AKDTree ($d_2$, two $c_i$ of $d_2$)\;
    }
    \uElseIf{$d$ is a slim cuboid}{
      get the counts $c_1,c_2$ from counts information\;
      split $d$ along the largest dimension to get $d_1$,$d_2$\;
      node.left\ \ = AKDTree ($d_1$, $c_1$)\;
      node.right = AKDTree ($d_2$, $c_2$)\;
    }
}
return node;
\end{algorithm}

To address the above high overhead issue of OpST, we propose an adaptive \kdtree, called \textbf{AKDTree}, to remove empty regions and extract sub-blocks (containing multiple unit blocks). AKDTree has a lower time complexity of $O(\frac{1}{3} N \cdot \log N)$ (will be discussed later). Figure~\ref{fig:kde} shows a simple 2D example. 
Specifically,
(1) we partition the data into small unit blocks. 
(2) We use a tree to hierarchically represent the whole data. Each node in the tree is associated with a sub-block of the data. 
Moreover, each node stores the number of non-empty unit blocks in the sub-block associated with the node. 
(3) For each node, we split its associated sub-block from the middle along one dimension to form two sub-blocks for its two children. Note that we select one dimension which can maximize the difference of the numbers of non-empty unit blocks of the two children (will be discussed in the next two paragraphs).
(4) We keep splitting a node until it has no empty unit block or itself is empty. 
(5) Once finishing the construction of the tree, we collect all the leaf nodes and send them to the compressor. Note that a non-empty leaf node does not have any empty unit block; otherwise, it will keep splitting. Thus, a leaf node must be an empty or full node, as shown in Figure~\ref{fig:kde}. 
The detailed algorithm is described in Algorithm~\ref{alg:kd}.

As mentioned in the step (3), we are distributing the non-empty unit blocks unevenly to two children for each node because we attempt to get as many leaf nodes with large sub-block sizes as possible. 
If we keep splitting sub-blocks in a fixed way, for instance, first split along the $x$-axis, second split along the $y$-axis, third split along the $x$-axis, fourth split along the $y$-axis, and so on, 
we will get a 2-by-2 sub-block for the node $n[2][2]$ as shown in the dashed box, while its largest possible sub-block could be 4 by 2.

\begin{figure}[t]
    \centering 
    \includegraphics[width=0.99\columnwidth]{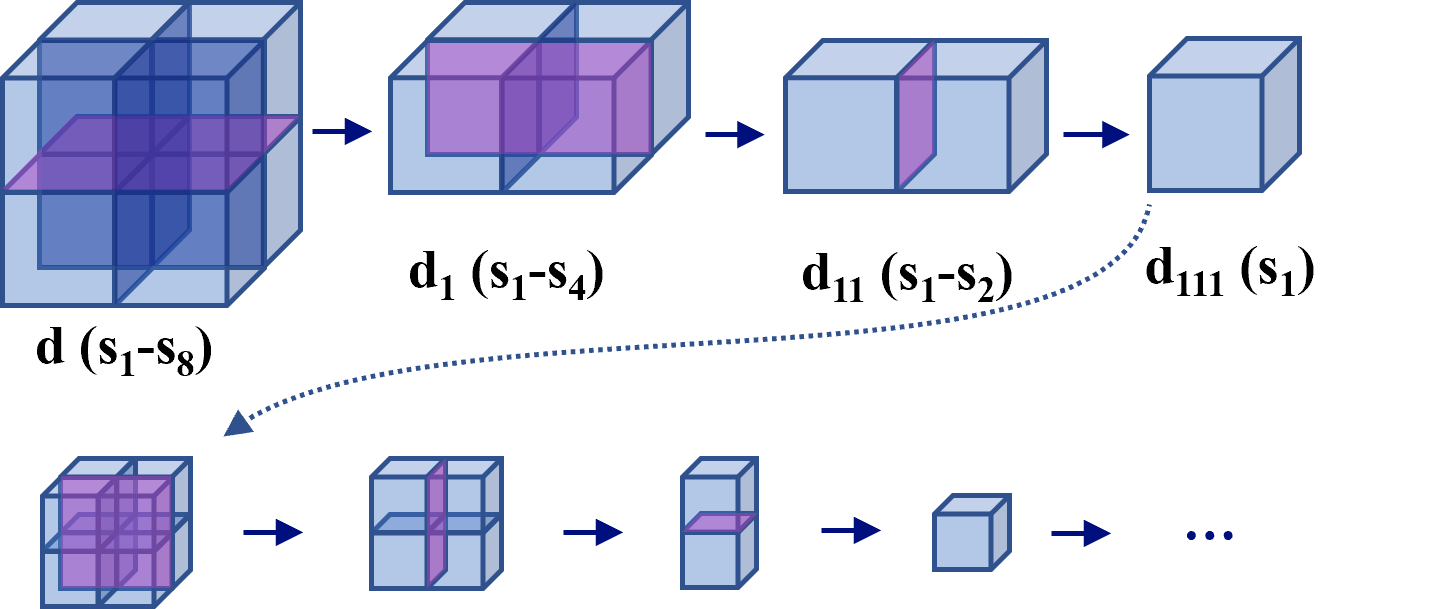}
    \vspace{-2mm}
    \caption{Example of the adaptive splitting, different shapes will have different number of choices for splitting. The process will be looped until a node is empty or full.}
    \label{fig:8sp}
\end{figure} 

To select one of the dimensions to unevenly distribute its non-empty unit blocks to the two children. 
We now present our dynamic splitting approach.
We categorize nodes into three different types: ``cube'' nodes, ``flat'' nodes, and ``slim'' nodes, whose dimension ratios are 1:1:1, 2:2:1, 2:1:1, respectively. 
First of all, for the cube node $d$, we first divide it into eight oct-blocks, i.e., $s_1$, $s_2$, $\cdots$, $s_8$ (as shown in Figure~\ref{fig:8sp}), each sized $\frac{n}{2}^3$. Here $n$ is the dimension size of the original data. 
Then, we can get the counts of non-empty unit blocks of the eight oct-blocks, i.e., $c_1$, $c_2$, $\cdots$, $c_8$. 
After that, We will decide along which dimension to split the cube node $d$ based on the counts. Specifically, we can calculate the following three difference values:
\begin{align*}
    \text{diff}_x &= |c_1+c_3+c_5+c_7-c_2-c_4-c_6-c_8|,\\
    \text{diff}_y &= |c_1+c_2+c_5+c_6-c_3-c_4-c_7-c_8|,\\
    \text{diff}_z &= |c_1+c_2+c_3+c_4-c_5-c_6-c_7-c_8|.
\end{align*}
Finally, we compare these three values and choose the dimension with the maximum difference to split. For example, if the maximum difference is $\text{diff}_z$, we will split $d$ along z-axis (i.e., the pink 2D plane shown in Figure~\ref{fig:8sp}) and get two flat nodes $d_1$ and $d_2$.

Then, for the flat nodes such as $d_1$, we can reuse $c_1$, $\cdots$, $c_4$ to decide whether to split $d_1$ along x-axis or y-axis by choosing the larger one among the following two difference values. 
\begin{align*}
    \text{diff}_x = |c_1+c_3-c_2-c_4|, \; \text{diff}_y = |c_1+c_2-c_3-c_4|.
\end{align*}
Finally, for the slim nodes such as $d_{11}$, 
we simply split it along x-axis to get two cube nodes $s_1$ and $s_2$. 
This process (i.e., cube nodes$\to$flat nodes$\to$slim nodes) in the step (3) will be looped until the node becomes to a leaf node (i.e., empty or full). 

Note that based on the above description, the counting process is required every three nodes in each three path (i.e., only for the ``cube'' nodes). 
Thanks to this dynamic splitting approach, we can lower the time complexity of the AKDTree algorithm to $O\big(\frac{1}{3} \cdot N \cdot \log N\big)$, where $N$ is the number of unit blocks, while extracting as many relatively large sub-blocks without empty unit block as possible.

In addition, after the dynamic splitting, we will have a series of sub-blocks with the same size but different directions (e.g., 2:2:1, 2:1:2, 1:2:2). We will align the sub-blocks with the same size based on their splitting dimensions (instead of transposing them in the memory), merge them into an array, and feed multiple merged arrays to the following compression. 


\subsection{Ghost-Shell Padding for High-density Data}\label{sec:gsp}

For high-density data such as z10's coarse level shown in  Figure~\ref{fig:dis_coarse} (i.e., about 77\% density), the benefit of using our proposed OpST or AKDTree is minimal because there is not much room for removing empty regions.
Meanwhile, due to the data partition/reorganization, OpST and AKDTree will hurt the data locality/smoothness.

\begin{figure}[h]
     \centering
     \begin{subfigure}[t]{0.49\linewidth}
         \centering
         \includegraphics[width=\linewidth]{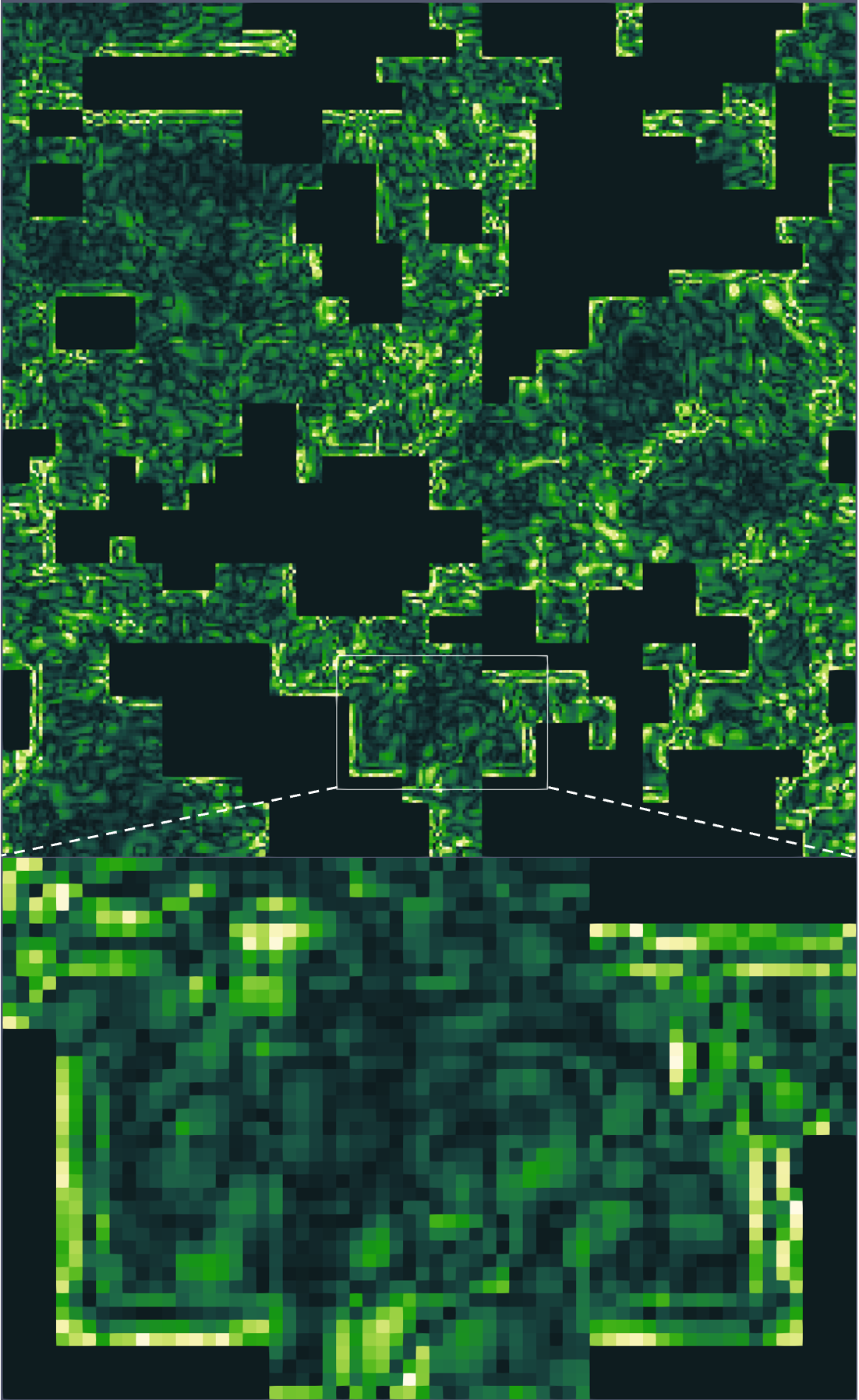}  
         \caption[t]{ZF (CR = 156.7, PSNR = 32.8 dB)}
         \label{fig:zero_err}
     \end{subfigure}
     \begin{subfigure}[t]{0.49\linewidth}
         \centering
         \includegraphics[width=\linewidth]{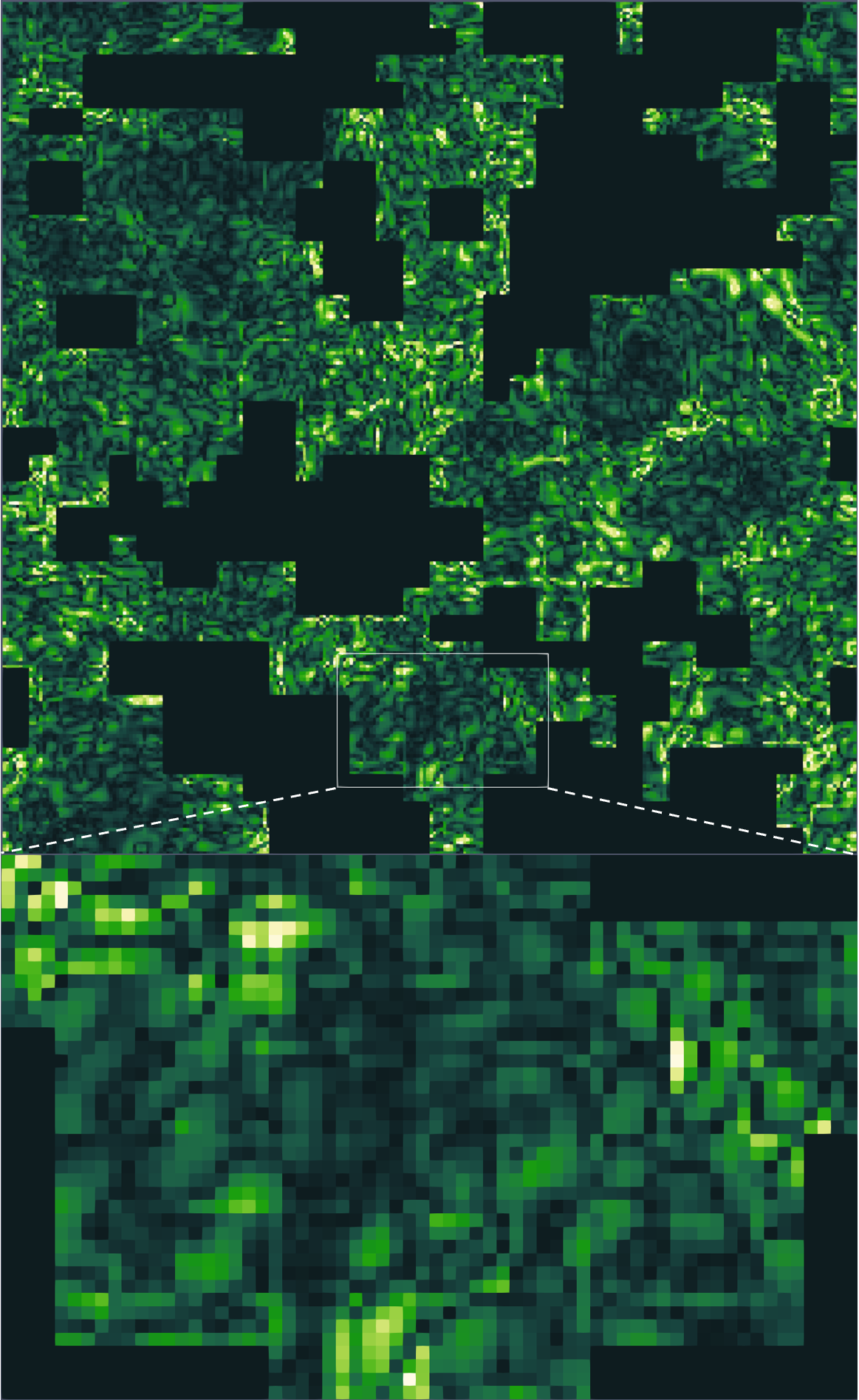}
         \caption{GSP (CR = 161.3, PSNR = 33.5 dB)}
         \label{fig:gsp_err}
     \end{subfigure}
     \vspace{-2mm}
        \caption[t]
        {Visual comparison (one slice) of compression errors of two approaches using SZ based on Nyx ``baryon density'' field (i.e., z10's coarse level, 77\% density). Brighter means higher compression error. The error bound is the relative error bound of $6.7\times10^{-3}$.}
        \label{fig:gsp_zero}
\end{figure}

To this end, we propose to pad zeros into the few empty regions, instead of removing them, followed by compression. 
However, these padded zeros can greatly reduce the performance of compression, especially for prediction-based lossy compression such as SZ, because these zeros can significantly affect the prediction accuracy of SZ, resulting in high compression errors on the boundaries, as shown in Figure~\ref{fig:zero_err}. 
More specifically, as mentioned in Section~\ref{sec:opst}, SZ uses each point's neighboring points' values to predict its value. 
Thus, for those boundary points which are adjacent to padded zeros, SZ will involve zero(s) into the prediction, while the actual values of these empty regions are typically non-zeros (saved in other AMR levels), which will seriously mislead the prediction. 

\begin{figure}[t]
    \centering 
    \includegraphics[width=0.90\columnwidth]{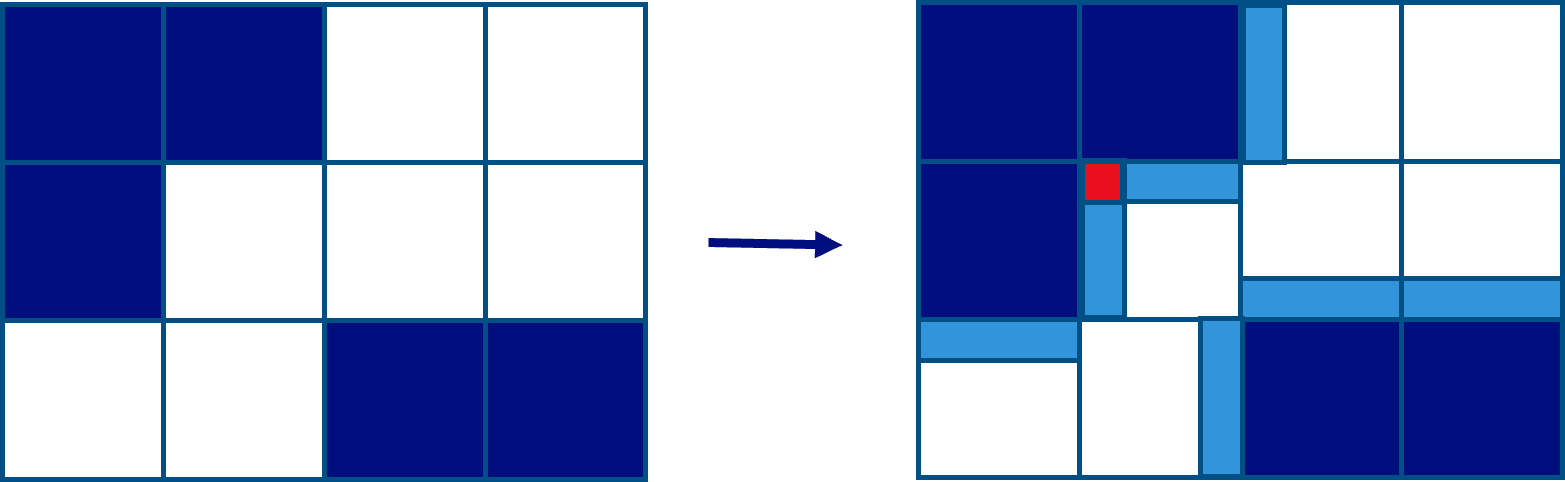}
     \vspace{-2mm}
    \caption{A 2D example of GSP approach. Non-empty blocks are in navy blue; padded blocks are in light blue/red; padded blocks based on more than one non-empty neighbors are in red. 
    }
    \label{fig:gsp-e}
\end{figure}



\begin{figure*}[h]
\centering
\begin{subfigure}{0.33\linewidth}\centering
    \includegraphics[width=0.99\linewidth]{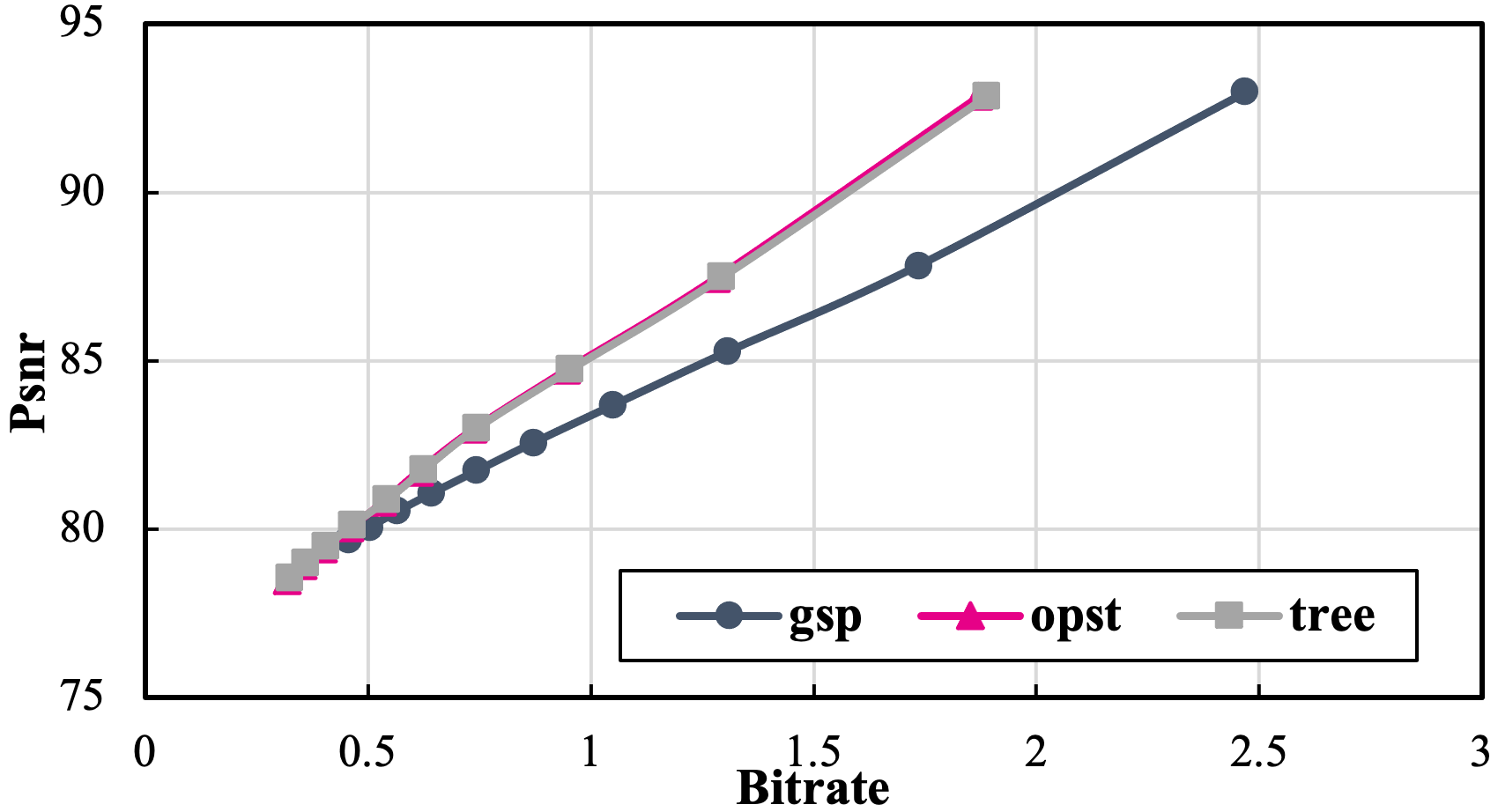}
    \vspace{-3mm}
	\caption{\footnotesize\centering Z10 (d = 23)}\label{fig:tt-1}
	\vspace{2mm}
\end{subfigure}
\begin{subfigure}{0.33\linewidth}\centering
    \includegraphics[width=0.99\linewidth]{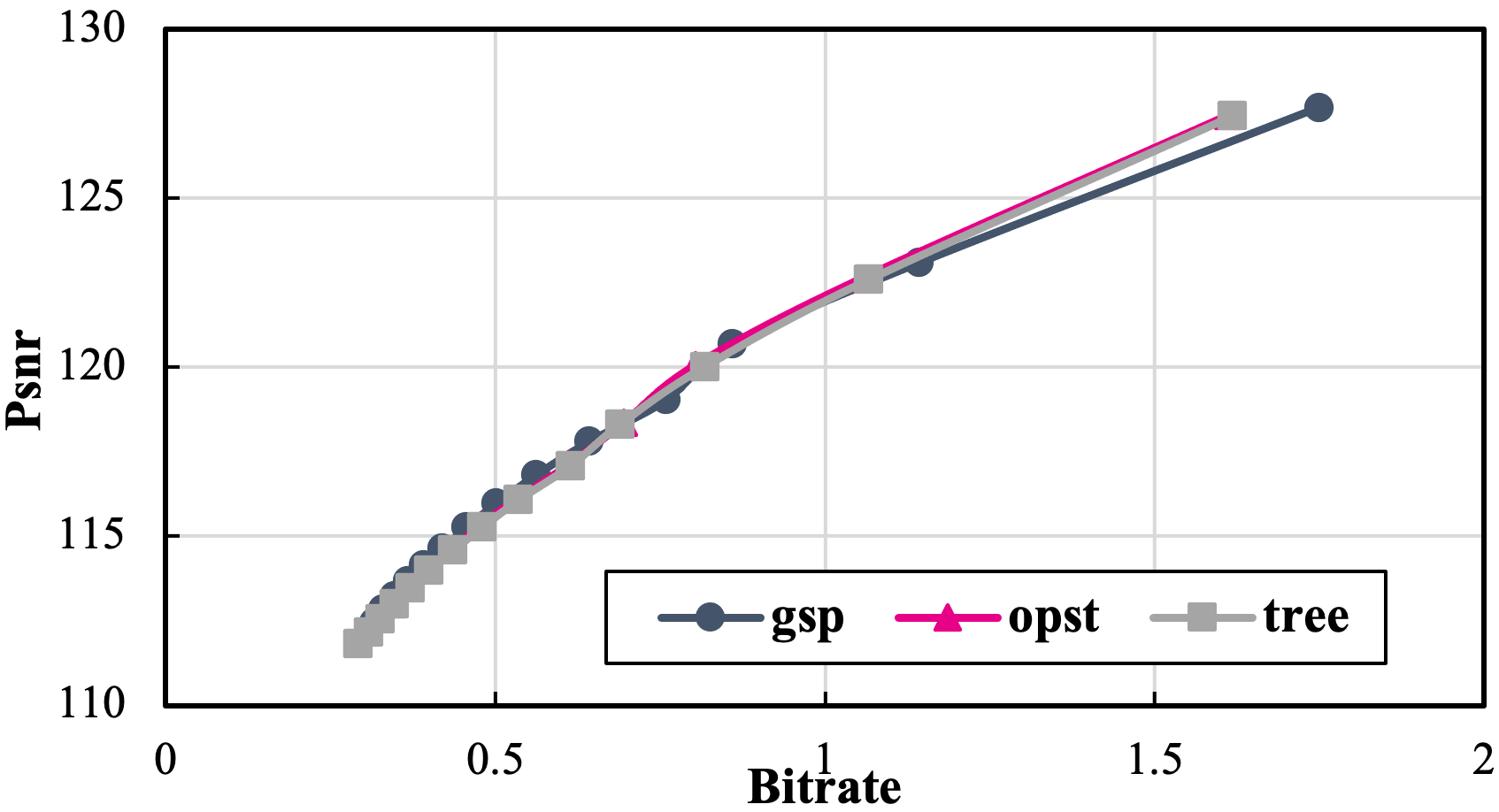}
    \vspace{-3mm}
	\caption{\footnotesize\centering z5 (d = 58)}\label{fig:tt-2}
	\vspace{2mm}
\end{subfigure}
\begin{subfigure}{0.33\linewidth}\centering
    \includegraphics[width=0.99\linewidth]{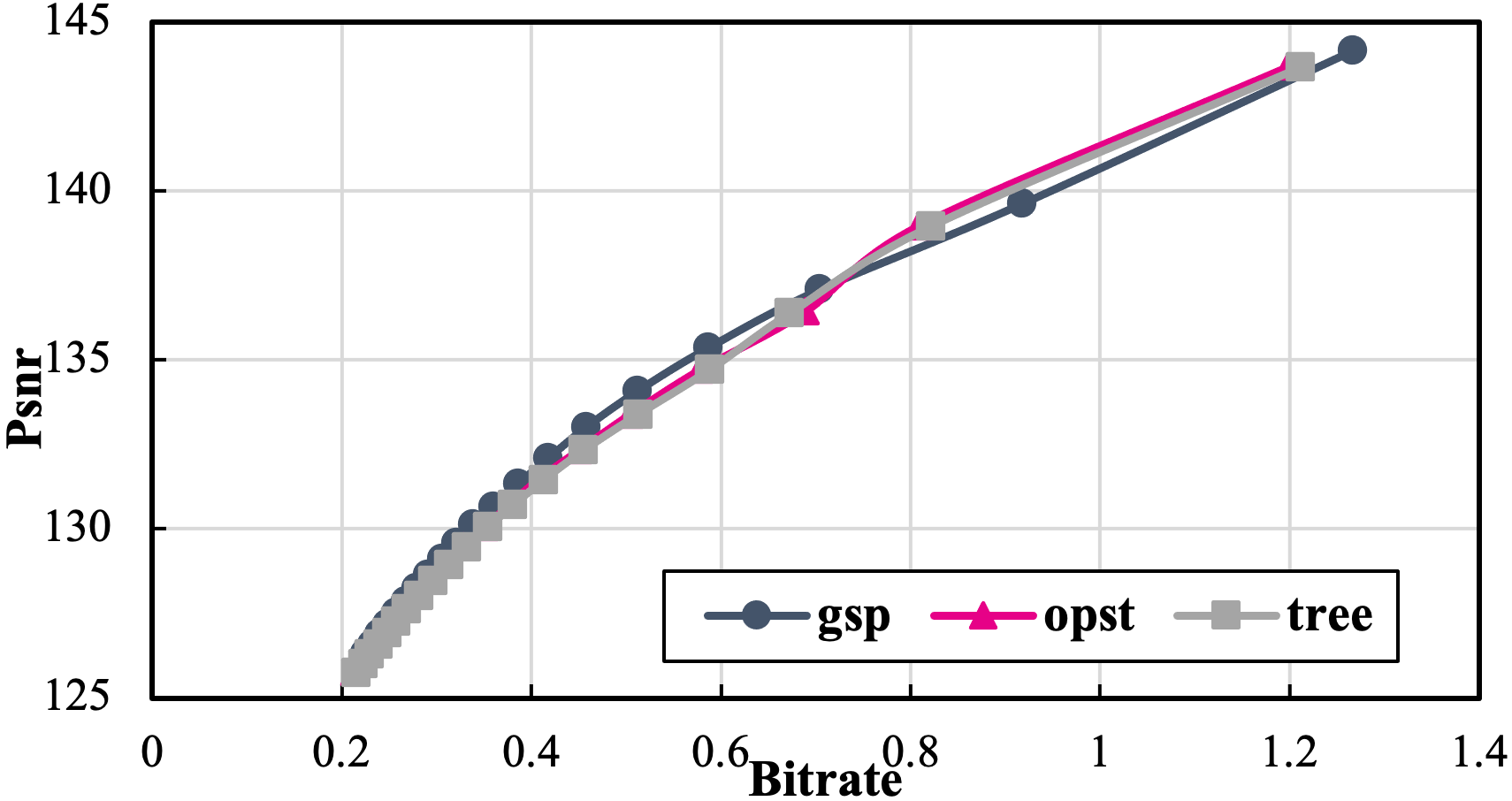}
    \vspace{-3mm}
	\caption{\footnotesize\centering z2 (d = 63)}\label{fig:tt-3}
	\vspace{2mm}
\end{subfigure}
\begin{subfigure}{0.33\linewidth}\centering
    \includegraphics[width=0.99\linewidth]{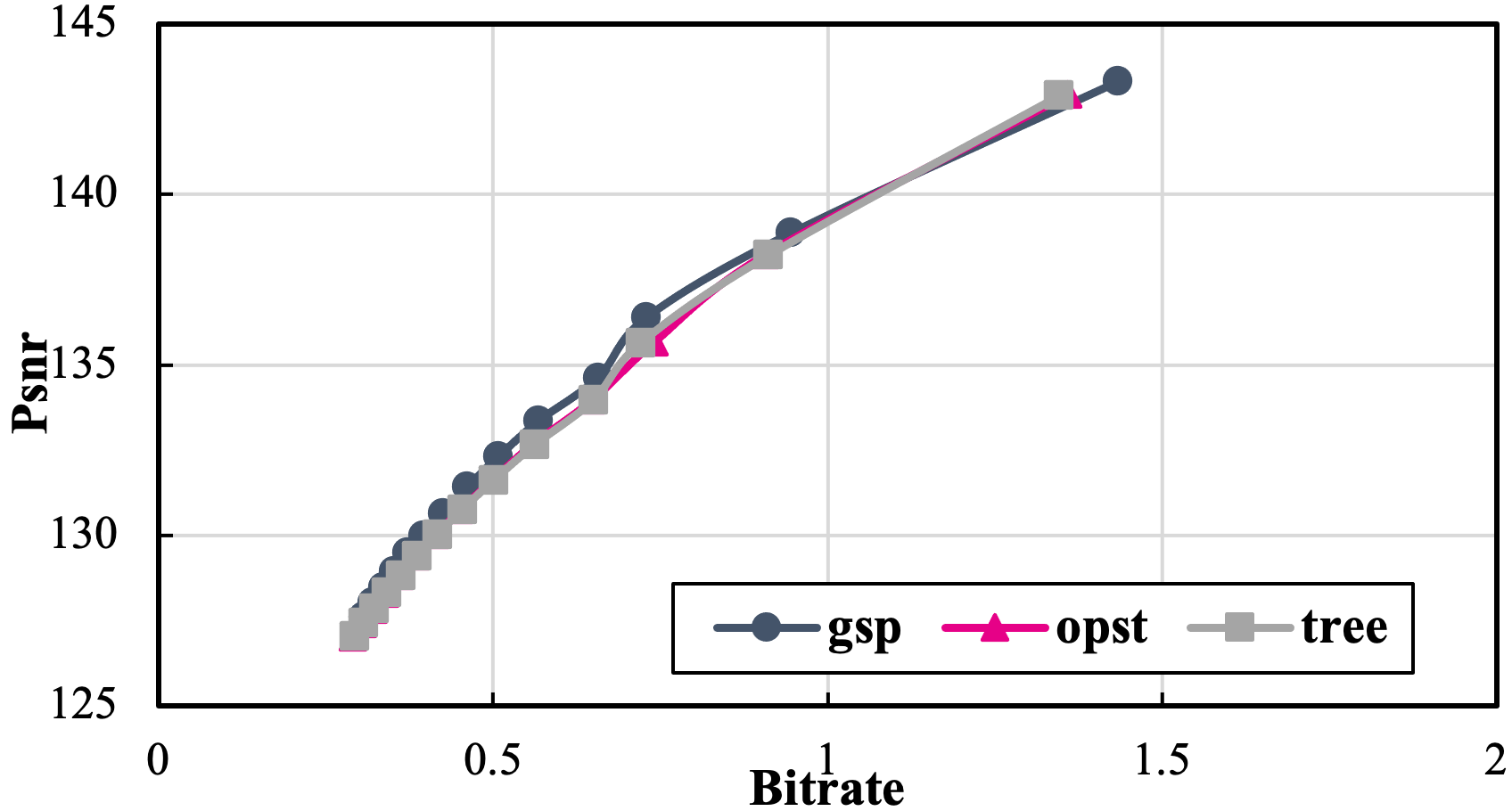}
    \vspace{-3mm}
	\caption{\footnotesize\centering Z3 (d = 64)}\label{fig:tt-4}
\end{subfigure}
\begin{subfigure}{0.33\linewidth}\centering
    \includegraphics[width=0.99\linewidth]{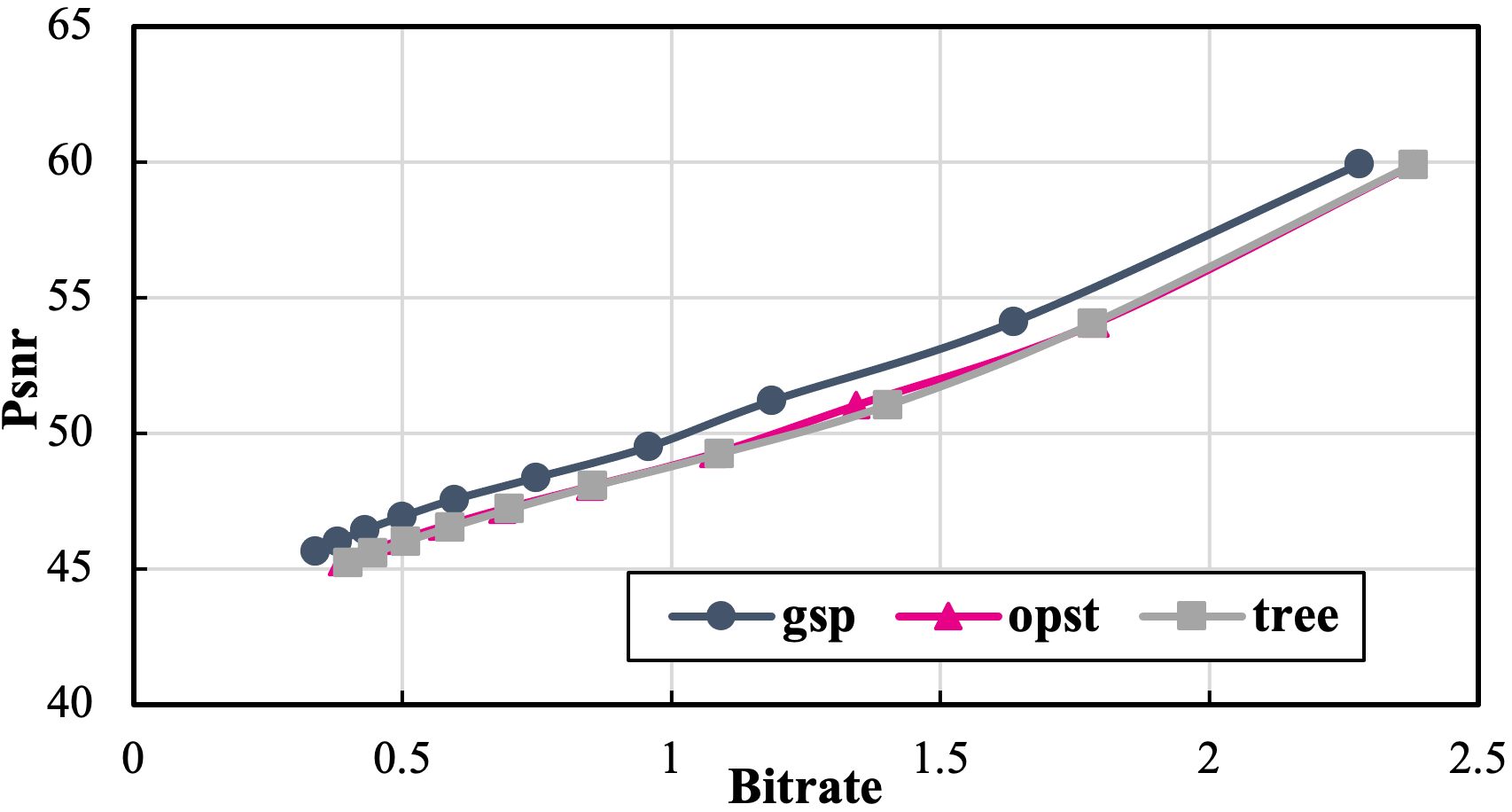}
    \vspace{-3mm}
	\caption{\footnotesize\centering d = 99.8}\label{fig:tt-5}
\end{subfigure}
\begin{subfigure}{0.33\linewidth}\centering
    \includegraphics[width=0.99\linewidth]{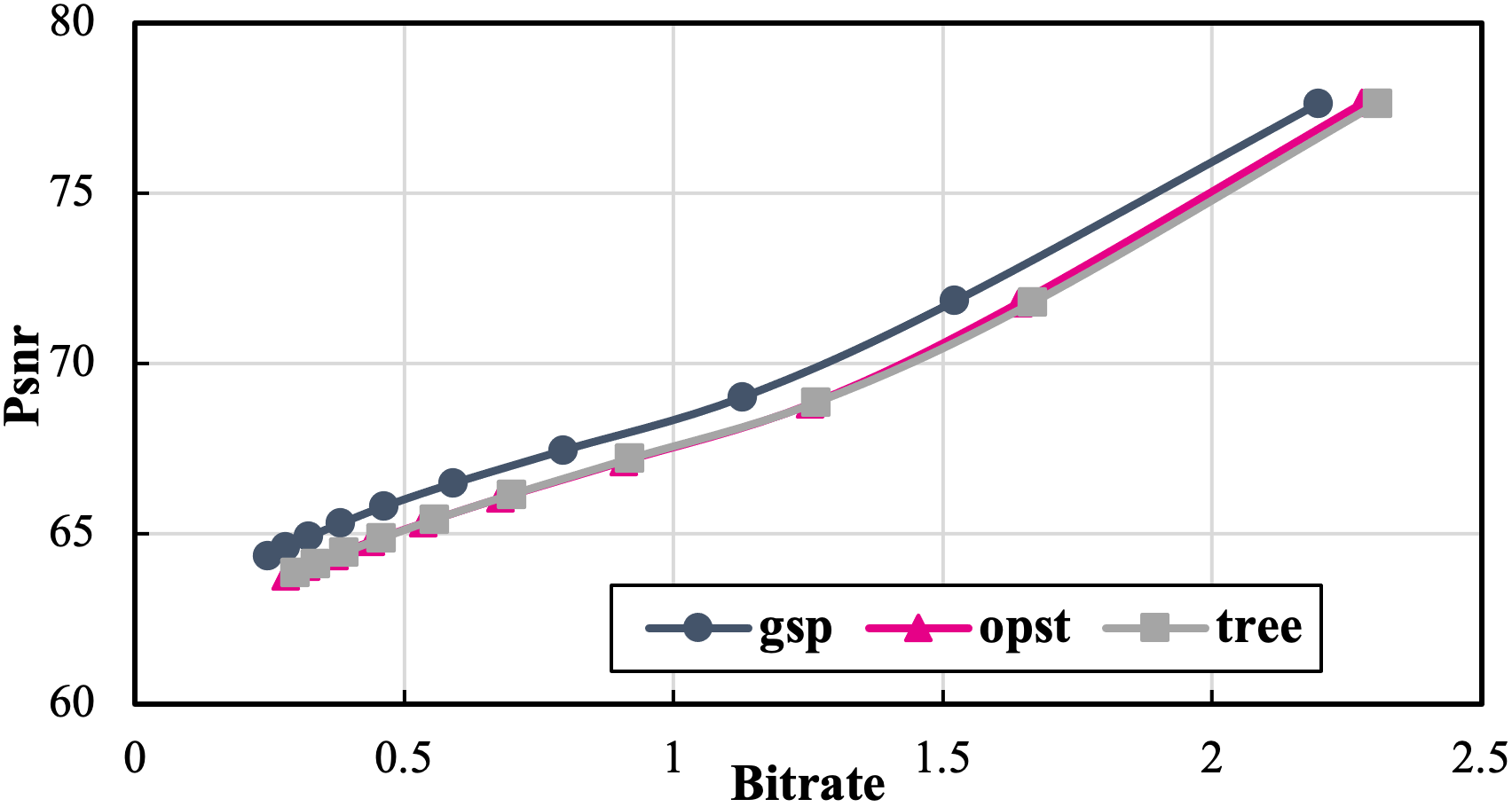}
    \vspace{-3mm}
	\caption{\footnotesize\centering d = 99.9}\label{fig:tt-6}
\end{subfigure}
\vspace{-2mm}
\caption{Compression performance comparison of GSP, OpST and AKDTree on six datasets with different densities.}
\label{fig:tt}
\vspace{-2mm}
\end{figure*}

To eliminate the above issue of padding zeroes, 
we propose to use a ghost-shell padding strategy (\textbf{GSP}) to \textcolor{black}{diffuse neighboring values to a padding layer.}
Figure~\ref{fig:gsp-e} illustrates the high-level idea, and the detailed algorithm is described in Algorithm~\ref{alg:gsp}. 
Specifically, we still partition the data into unit blocks. 
\textcolor{black}{Then, we will pad each empty unit block by using the average of its non-empty neighbors' boundary data values.}
Note that some empty unit blocks can have more than one non-empty neighbors such as the red box shown in Figure~\ref{fig:gsp-e}. For these blocks, we will use the average value of all its neighbors for padding. 
Correspondingly, we will remove these padded values during the decompression based on the saved padding information. 
Note that since the padding process is only for non-empty blocks, this metadata overhead is almost negligible for high-density data (e.g.,  0.1\%).

After padding, each boundary point will be predicted using the average of all the boundary data in the unit block(s) to which it belongs or is neighbored. 
As shown in Figure~\ref{fig:gsp_zero}, compared to the zero filling (ZF) approach, GSP can significantly reduce the overall compression error, especially for the boundary data. 
Moreover, the GSP approach can provide a similar compression ratio to the ZF approach on this high-density data and hence a better rate-distortion. 
A detailed evaluation will be presented in Section~\ref{sec:evaluation}.

\begin{algorithm}[h]
\caption{Proposed Ghost Shell Padding Method}\label{alg:gsp}  
\KwIn {Data, $x$, $y$}
\KwOut {Data after padding}
\For{each unit block $b_i$}{
  \If{$b_i$ is empty \AND $b_i$ has non-empty neighbor }{
    \For{each non-empty neighbor $n_j$}{
        pad slice = avg (first $y$ slices of $n_j$ next to $b_i$)\;
      \uIf{overlap edge}{
      $pad = {pad}/{2}$\;
      }
      \uElseIf{overlap corner}{
      $pad = {pad}/{3}$\;
      }
      \Else{
      continue\;
      }
      add an $x$-layers pad slice to $b_i$ next to $n_j$\;
    }
  }
}
return padded Data
\end{algorithm}

\subsection{Hybrid Compression Strategy} \label{sec:hybrid}

In this section, we propose a solution to adaptively choose a best-fit compression strategy from on our proposed \textit{OpST}, \textit{AKDTree}, and \textit{GSP} based on the data characteristics (i.e., data density).
According to Section~\ref{sec:opst},~\ref{sec:kd}, and~\ref{sec:gsp}, the OpST approach is more suitable for sparse (i.e., low-density) data, while the AKDTree approach is designed to address the high time overhead of OpST when the density of data increases. 
When the data density is very high, the GSP approach will be used to maintain the data smoothness/locality compared to the AKDTree and OpST approaches. 
Therefore, we propose to use two data-density thresholds to determine when to use OpST, AKDTree, or GSP.  

\begin{figure}[h]
    \centering 
    \includegraphics[width=0.95\columnwidth]{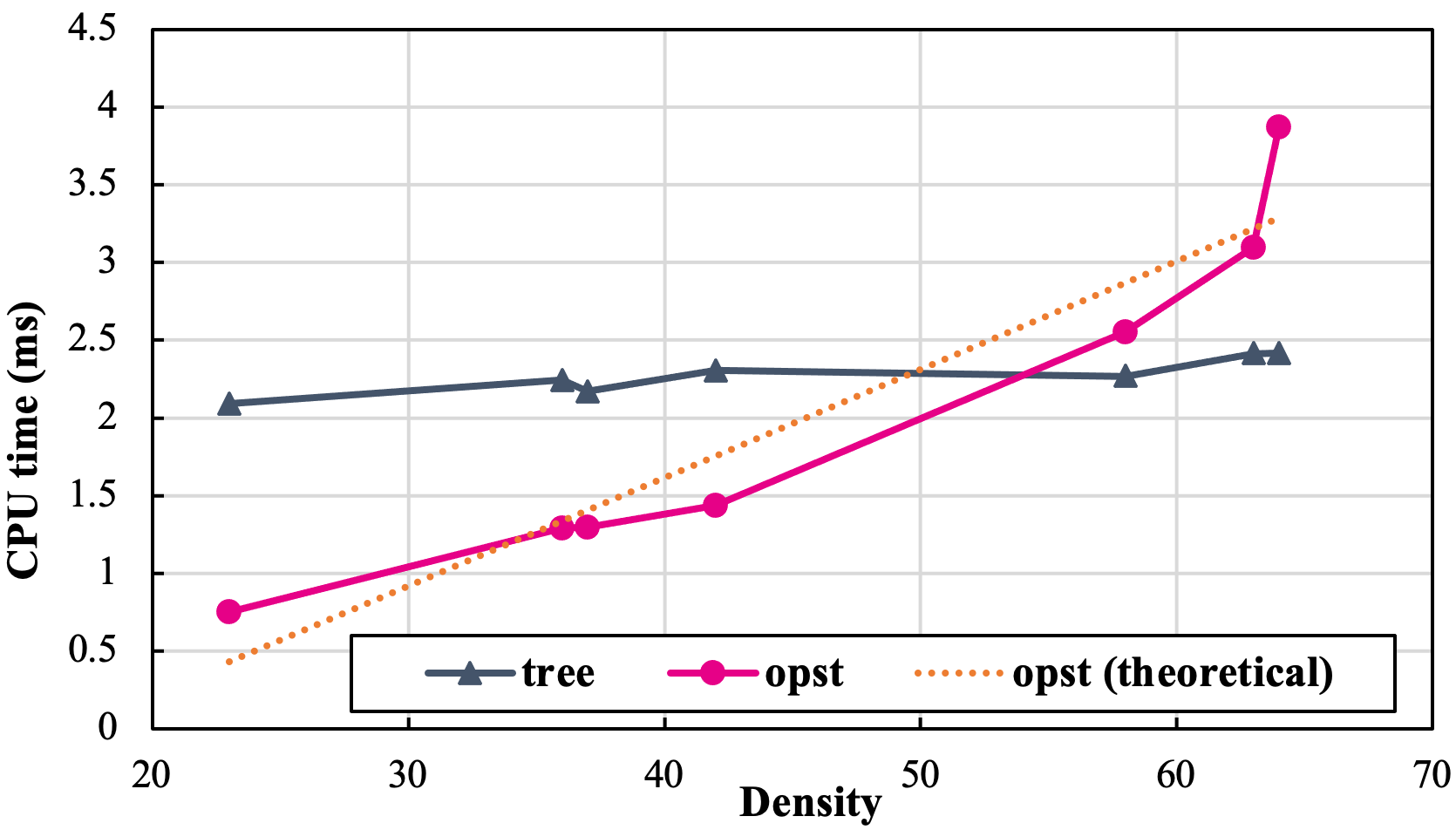}
    \vspace{-2mm}
    \caption{Time overhead comparison 
    of OpST and AKDTree on different datasets with different densities.}
    \label{fig:time}
\end{figure} 

To decide the first threshold $T_1$ for switching between OpST and AKDTree, we perform a series of experiments, as shown in Figure~\ref{fig:tt}. 
The figure shows that OpST and AKDTree have almost identical compression performance in terms of bit-rate and PSNR on all six datasets/levels (from different timesteps) with different densities. 
Moreover, Figure~\ref{fig:time} shows the time costs of OpST and AKDTree (excluding compression). The figure demonstrates that the time of AKDTree is relatively stable, while the time of OpST increases linearly with the increase of data density. 
Overall, the only criterion for selecting OpST or AKDTree is the time cost rather than the compression performance. 
This is consistent with our previous design aim, that is, AKDTree is mainly designed to address the high time overhead issue of OpST. 
Since OpST and AKDTree have a similar speed when the density is around 50\%, we propose to choose $T_1$ = 50 for choosing OpST or AKDTree.


Next, to determine the threshold $T_2$ for switching between AKDTree and GSP, we also evaluate them on different datasets with different densities. 
As shown in Figure~\ref{fig:tt}, when the density is relatively low, AKDTree outperforms GSP with respect to both bit-rate and PSNR; when the density gets higher and higher, GSP gradually outperforms AKDTree. 
We can also observe that AKDTree and GSP have similar compression performance when the density is around 60\%. Thus, we use $T_2$ = 60\% for choosing AKDTree or GSP. 

In summary, our proposed hybrid compression approach is described as follows.
\begin{enumerate}[topsep=0pt,partopsep=1ex,parsep=0pt]
    \item When the density is smaller than $T_1 = 50\%$, we will use OpST to remove empty regions and then perform the compression;
    \item When the density is between $T_1 = 50\%$ and $T_1 = 60\%$, we will use AKDTree to remove empty regions and then compress;
    \item When the density is larger $T_1 = 60\%$, we will use GSP to pad appropriate values and then compress the padded data.
\end{enumerate}


%% file: tex/04_evaluation.tex
\section{Experimental Evaluation}
\label{sec:evaluation}

\textcolor{black}{In this section, we first present our experimental setup and evaluation metrics. We then demonstrate and discuss the effectiveness of \textsc{TAC} in terms of both compression ratio and data quality. After that, we show the benefit of using adaptive error bound in \textsc{TAC} regarding post-analysis quality. Finally, we show that \textsc{TAC} has comparable throughput compared to comparison baselines.}

\subsection{Experimental Setup}
\textit{Test data.} Our evaluation mainly focuses on the AMReX framework~\cite{zhang2019amrex}, particularly the Nyx cosmology simulation~\cite{nyx}. Nyx is a state-of-the-art extreme-scale cosmology code using AMReX, which generates six fields including baryon density, dark matter density, temperature, and velocities ($x$, $y$, and $z$). 
We use \textcolor{black}{seven} datasets generated by two real-world simulation runs with different numbers of AMR levels, simulating a region of 64 megaparsecs (Mpc). For this data, Z is equal to the redshift, i.e. the displacement distant galaxies and celestial objects, as seen in Tab~\ref{tab:datasets1}.

Specifically, 
the first run has \textcolor{black}{two} levels of refinement, with the coarse level of $256^3$ grids and the fine level of $512^3$ grids. 
We've collected \textcolor{black}{five} timesteps with the finest level density from \textcolor{black}{23\% to 64\%}.
The second run has a maximum of four levels of refinement. It was initially configured at the resolution of $128^3$ and gradually refined to $1024^3$. This run collected \textcolor{black}{three} timesteps with the coarsest-level resolution of $256^3$ (two levels), $512^3$ (three levels), and finest $1024^3$ (four levels), respectively. The density of the finest level varies from \textcolor{black}{0.2\% to 0.003\%}.
Note that the density of the finest level describes how much of the data in the dataset is at the highest resolution; a higher density of the finest level means that more data is refined to the highest resolution. Usually, the data density is gradually increasing at the finest level, within a single run. 




\textit{Evaluation platform.}
The test platform is equipped two 28-core Intel Xeon Gold 6238R processors and 384 GB DDR4 memory. 

\begin{table}[ht]
\caption{Our tested datasets.}
\vspace{-2mm}
\resizebox{\linewidth}{!}{%
\input{Table/dataset}
 }
  \label{tab:dataset}%
\end{table}%

\textit{Comparison baselines.}
As discussed in Section~\ref{sec:background}, we have three 1D or 3D comparison baselines.
Specifically, (1) the \textit{1D baseline (naive)}: each AMR level is compressed separately as a 1D array; (2) the \textit{1D baseline (zMesh)}~\cite{zMesh}: we refer readers to Section~\ref{sec:background} for more details about how the zMesh approach reorganize the AMR data for 1D compression; and (3) the \textit{3D baseline}: Different AMR levels are unified to the same resolution for 3D compression.

\begin{figure*}[h]
     \centering
     \begin{subfigure}[t]{\columnwidth}
         \centering
         \includegraphics[width=0.98\columnwidth]{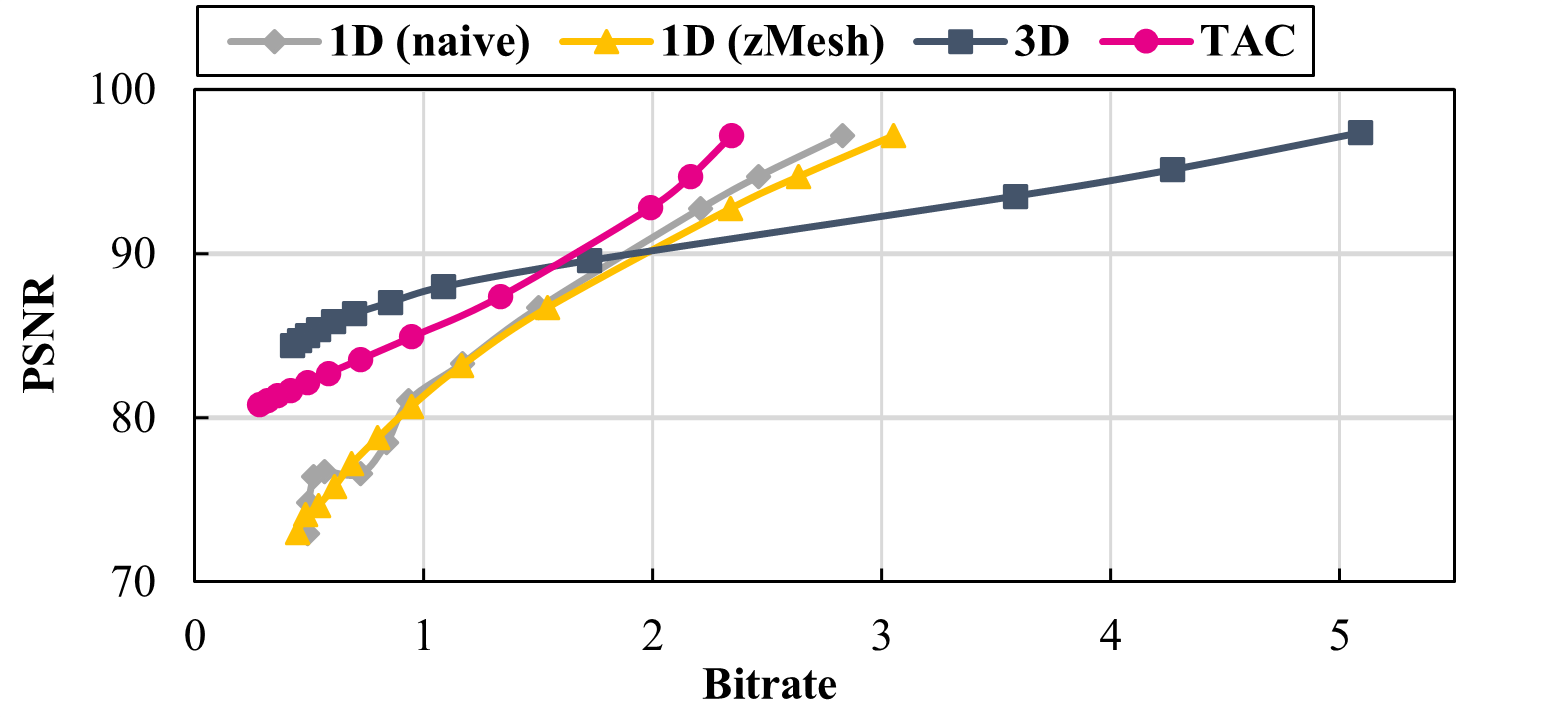} 
         \vspace{-2mm}
         \caption[t]{Run1\_Z10 (finest-level density = 23\%)}
         \label{fig:z10}
     \end{subfigure}
     \begin{subfigure}[t]{\columnwidth}
         \centering
         \includegraphics[width=0.98\columnwidth]{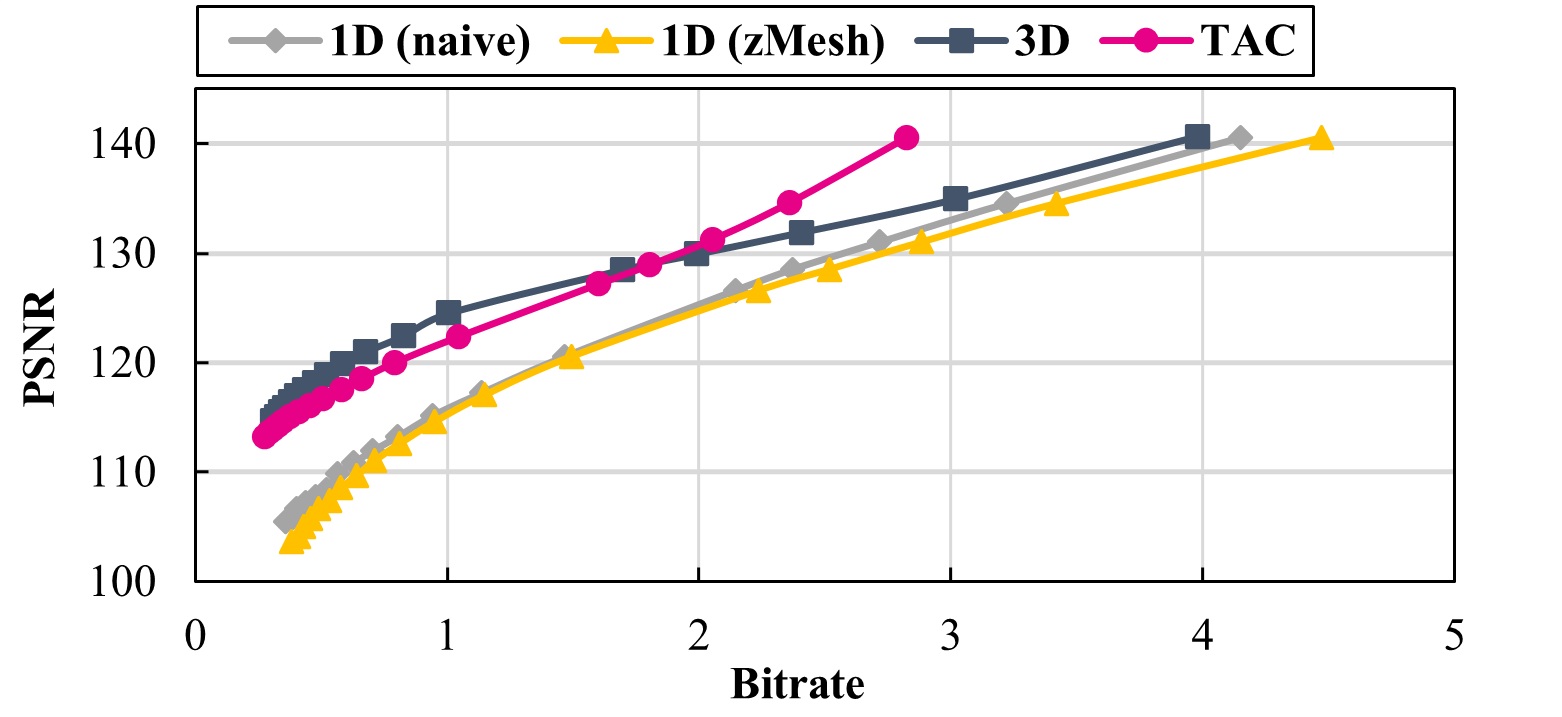}
         \vspace{-2mm}
         \caption{Run1\_Z5 (finest-level density = 58\%)}
         \label{fig:z5}
     \end{subfigure}
     \begin{subfigure}[t]{\columnwidth}
         \centering
         \includegraphics[width=0.98\columnwidth]{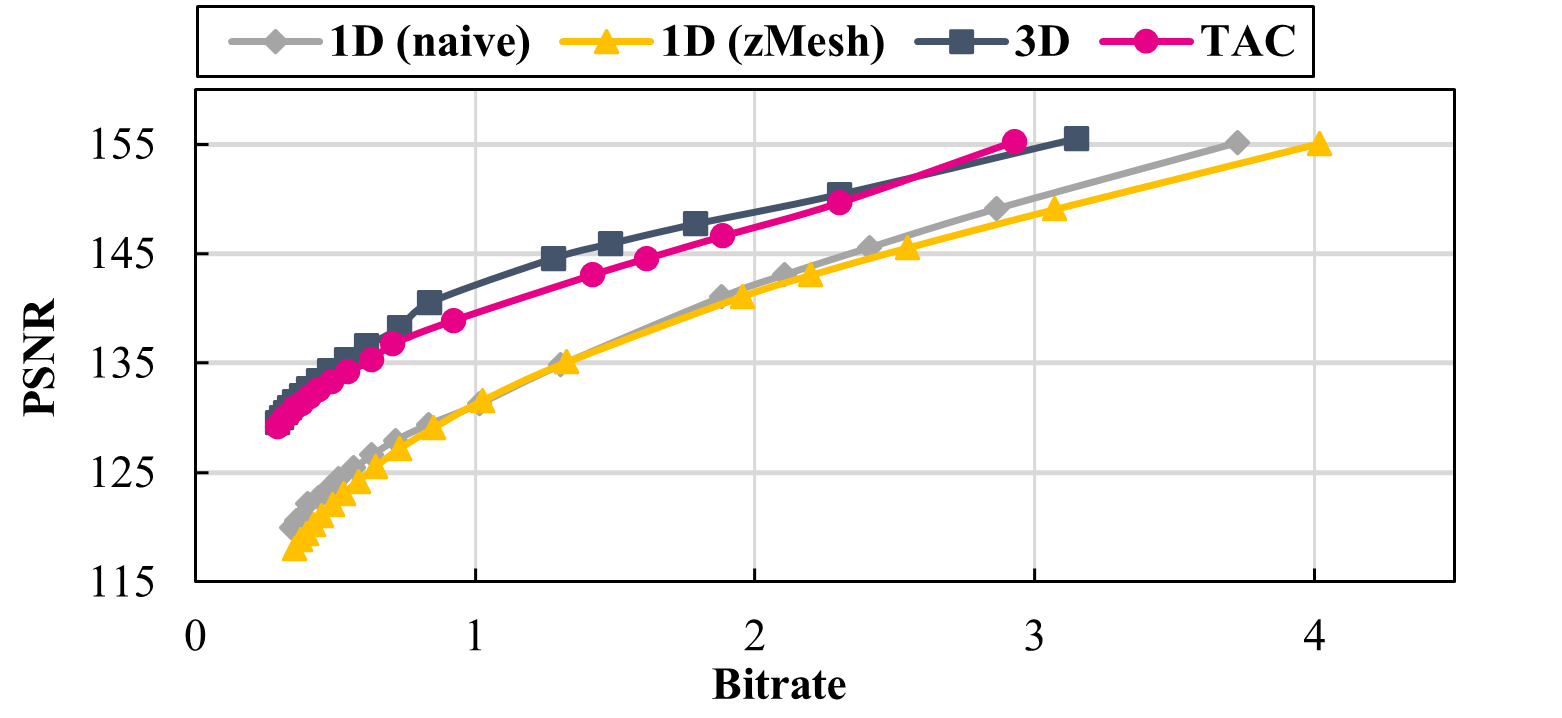}
         \vspace{-2mm}
         \caption{Run1\_Z3 (finest-level density = 64\%)}
         \label{fig:z3}
     \end{subfigure}
     \begin{subfigure}[t]{\columnwidth}
         \centering
         \includegraphics[width=0.98\columnwidth]{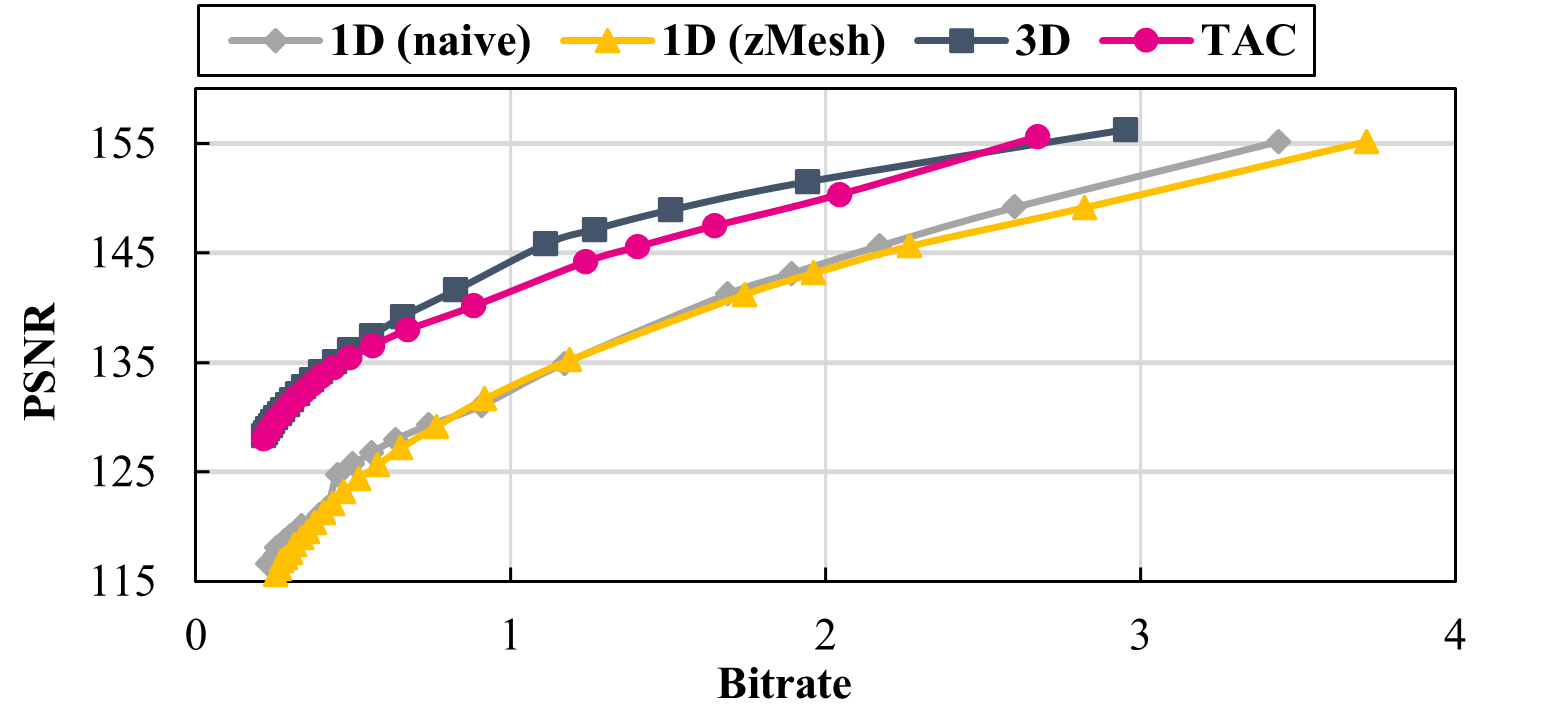}
         \vspace{-2mm}
         \caption{Run1\_Z2 (finest-level density = 63\%)}
         \label{fig:z2}
     \end{subfigure}
     \vspace{-2mm}
        \caption[t]{Rate-distortion comparison of \textsc{TAC} and baselines on the early time-step (Z10) to the late time-step (Z2) from run1.
        }
        \vspace{-4mm}
        \label{fig:vsbs_1}
\end{figure*}

\subsection{Evaluation Metrics} \label{metric}
We will evaluate the compression performance based on the following metrics including generic and application-specific metrics.

\begin{enumerate}
    \item Compression ratio or bit-rate (generic, Section \ref{subsec:evaRateDis})
    \item Distortion quality (generic, Section \ref{subsec:evaRateDis})
    \item Compression throughput (generic, Section \ref{subsec:throughput})
    \item Rate-distortion (generic, Section \ref{subsec:evaRateDis})
    \item Power spectrum (cosmology specific, Section \ref{subsec:post-analysis})
    \item Halo finder (cosmology specific, Section \ref{subsec:post-analysis})
\end{enumerate}

\textbf{Metric 1:} To evaluate the size reduction as a result of the compression, we use the compression ratio, defined as the ratio of the original data size compared to the compressed data size, or bit-rate (bits/value), representing the amortized storage cost of each value. For a single-/double-precision floating-point data, the bit-rate is 32/64 bits per value before compression. 
The compression ratio and bit-rate has a mathematical relationship as their product is 32/64 so that a lower bit-rate means a higher compression ratio.

\textbf{Metric 2:} Distortion is another important metric used to evaluate lossy compression quality in general. We use the peak signal-to-noise ratio (PSNR) to measure the distortion quality.
\begin{equation*}
\textstyle
\text{PSNR} = 20\cdot \log_{10}\left(R_X\right) - 10\cdot \log_{10}\left(\sum_{i=1}^{N} {e_i^2}/{N}\right),
\end{equation*}
where $e_i$ is the difference between the original and decompressed values for the point $i$, $N$ is the number of points, and $R_X$ is the value range of the dataset $X$. Note that higher PSNR less error.

\textbf{Metric 3:} Similar to prior work~\cite{sz17,sz18,liang2018efficient,liang2021error,jin2020understanding,jin2021adaptive,zhao2020significantly}, we plot the rate-distortion curve to compare the distortion quality with the same bit-rate, for a fair comparison between different compression approaches, taking into account diverse compression algorithms.

\textbf{Metric 4:} (De)compression throughputs are critical to improving the I/O performance. We will calculate the throughput based on the original data size and (de)compression time. 

\textbf{Metric 5:} 
Matter distribution in the Universe has evolved to form astrophysical structures on different physical scales, from planets to larger structures such as superclusters and galaxy filaments. The two-point correlation function $\xi(r)$, which gives the excess probability of finding a galaxy at a certain distance $r$ from another galaxy, statistically describes the amount of the Universe at each physical scale. The Fourier transform of $\xi(r)$ is called the matter power spectrum $P(k)$, where $k$ is the comoving wavenumber. 
The matter power spectrum describes how much structure exists at each physical scales. 
We run power spectrum on the baryon density field by using a cosmology analysis tool called Gimlet. We compare the power spectrum $p'(k)$ of decompressed data with the original $p(k)$ and accept a maximum relative error within 1\% for all k < 10.

\textbf{Metric 6:} 
Halo finder aims to find the halos (over-densities) in the dark matter distribution and output the positions, the number of cells, and mass for each halo it finds, respectively. Specifically, the halo-finder algorithm~\cite{Davis1985} searches for the halos from all the simulated data, with the following two criteria: (1) the mass of a data point must be greater than a threshold (e.g., 81.66 times the average mass of the whole dataset) to become a halo cell candidate~\cite{jin2020understanding, jin2021adaptive, ffis}, and (2) there must be enough halo cell candidates in a certain area to form a halo. For decompressed data, some of the information (mass and cells of halos) can be distorted from the original. 









\begin{figure}[t]
     \centering
     \begin{subfigure}[t]{\columnwidth}
         \centering
         \includegraphics[width=0.98\columnwidth]{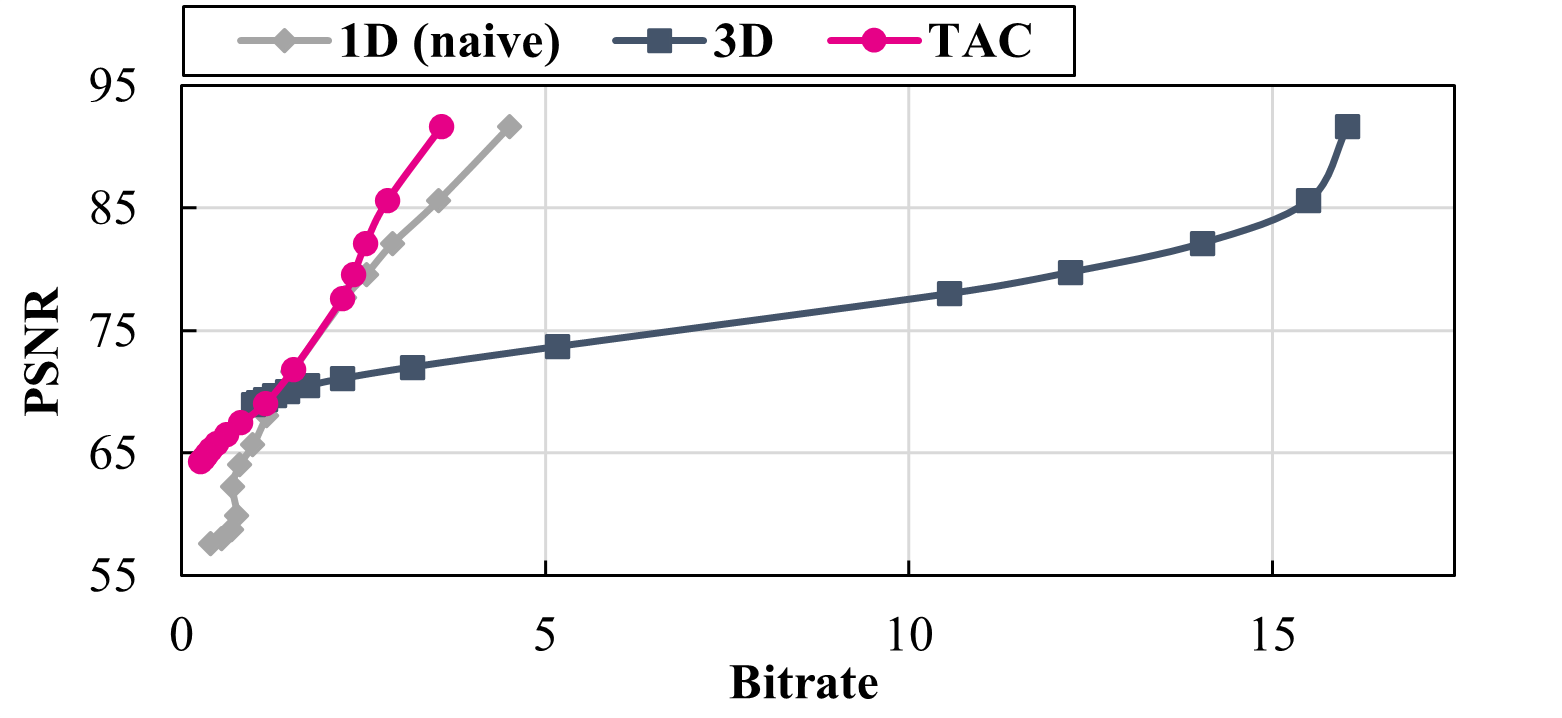}  
         \vspace{-2mm}
         \caption{Run2\_T2 (finest-level density = 0.2\%)}
         \vspace{2mm}
         \label{fig:t2}
     \end{subfigure}
     \begin{subfigure}[t]{\columnwidth}
         \centering
         \includegraphics[width=0.98\columnwidth]{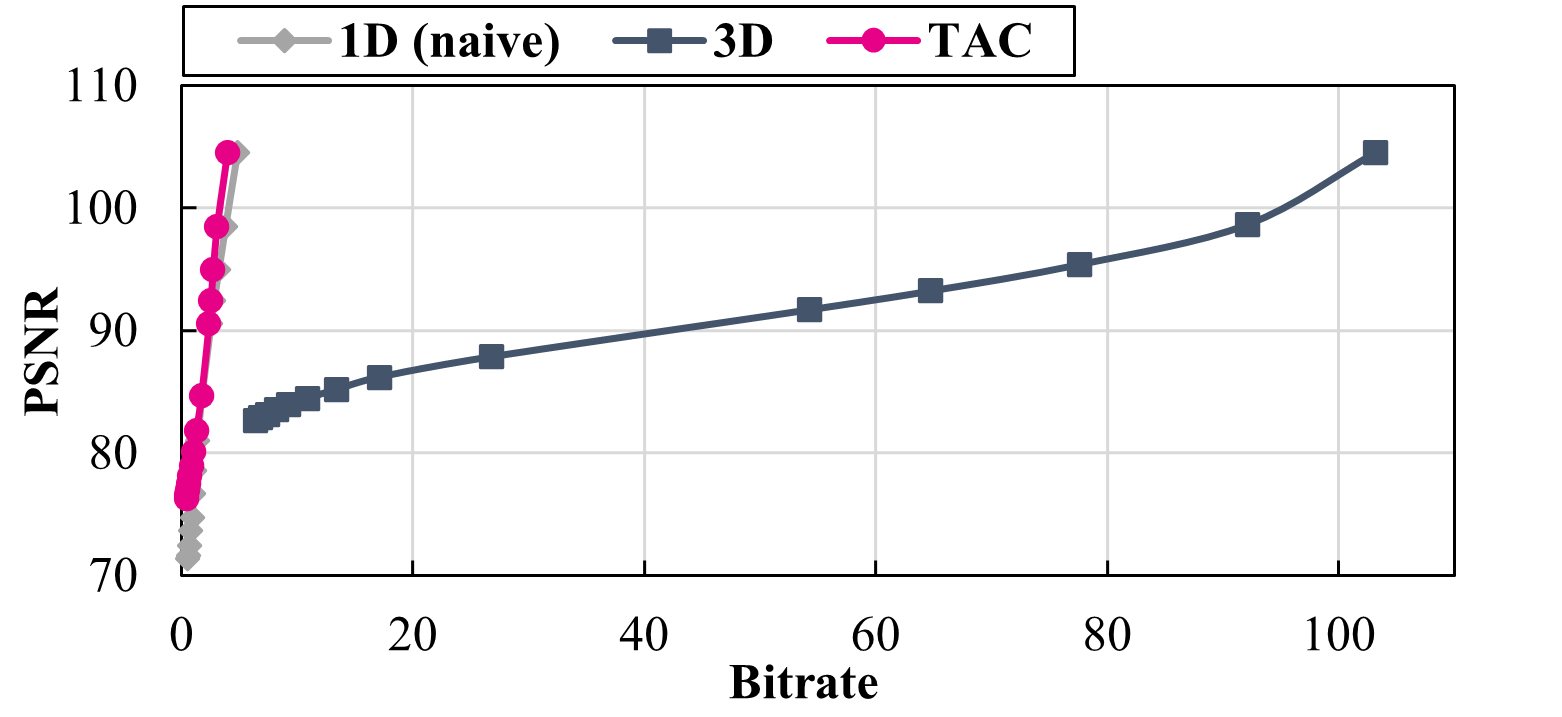}
         \vspace{-2mm}
         \caption{Run2\_T3 (finest-level density = 0.02\%)}
         \vspace{2mm}
         \label{fig:t3}
     \end{subfigure}
     \begin{subfigure}[t]{\columnwidth}
         \centering
         \includegraphics[width=0.98\columnwidth]{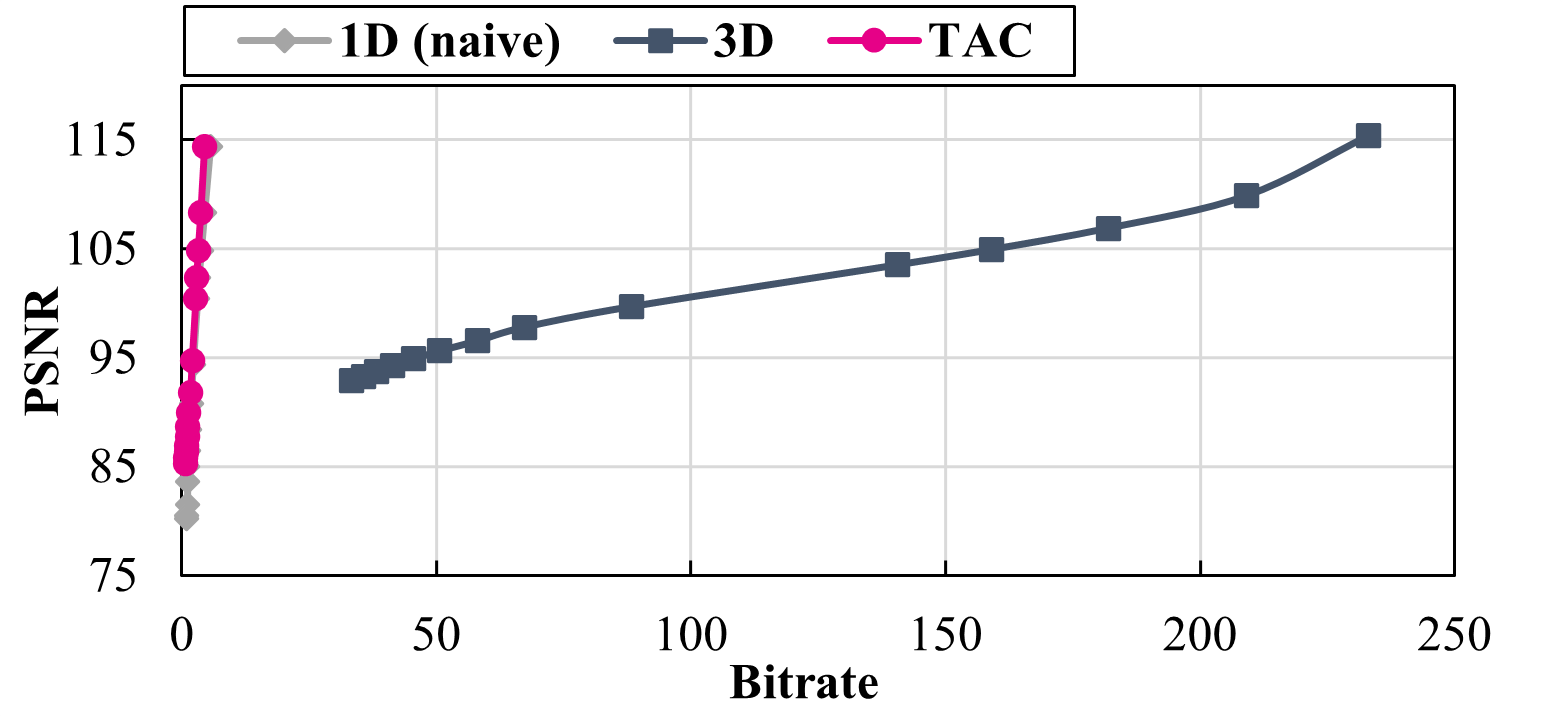}
         \vspace{-2mm}
         \caption{Run2\_T4 (finest-level density = 3E-5)}
         \label{fig:t4}
     \end{subfigure}
     \vspace{-4mm}
        \caption[t]{Rate-distortion comparison of \textsc{TAC} (top-left) and baselines on different time-steps from run2.
        }
        \vspace{-2mm}
        \label{fig:vsbs_2}
\end{figure}

\subsection{Evaluation on Rate-distortion} 
\label{subsec:evaRateDis}
We first evaluate the rate-distortion of \textsc{TAC} and compare it with the baselines 
on different datasets. 

For the 1D baseline, as shown in Figure~\ref{fig:vsbs_1} and~\ref{fig:vsbs_2}, \textsc{TAC} (top-left curve) outperforms the 1D baseline across all the 7 datasets. Furthermore, the performance of \textsc{TAC} is more stable (i.e., smoother curve) than the 1D baseline. We can also find that zMesh is slightly worse than the 1D baseline on our tested data, which will be explained in the next section.

For the 3D baseline, we can observe that \textsc{TAC} has much better performance when the finest level has a relatively low density or the decompressed data has a high PSNR, as shown in Figure~\ref{fig:vsbs_2}. However, when the finest level has a relatively high density, \textsc{TAC} cannot dominate the 3D baseline as shown in Figure~\ref{fig:z3} and~\ref{fig:z2}. 
Specifically, in Figure~\ref{fig:z10} (the finest level density is 23\%), \textsc{TAC} outperforms the 1D baseline when the bit-rate is larger than 1.6; in Figure~\ref{fig:z5} (the finest level density is 58\%), the intersection is the bit-rate of 1.9; as the finest level density continues to grow up to 63 and 64 in Figure~\ref{fig:z3} and~\ref{fig:z2}, \textsc{TAC} is slightly worse than the 3D baseline until the bit-rate is larger than 2.5.
In the next section, we will discuss why the 3D baseline is slightly better in the datasets of which finest level has a very high density in detail later and will also propose a solution to adaptively use the 3D baseline and \textsc{TAC}.



\subsection{Discussion on Comparison with Baselines} \label{sec:dis}

On compression, zMesh is meant to improve the smoothness of the block-structured AMR datasets by taking advantage of the data redundancy between each AMR level (as described in the introduction). 

\begin{figure}[h]
     \centering
     \begin{subfigure}[t]{0.45\linewidth}
         \centering
         \includegraphics[width=\linewidth]{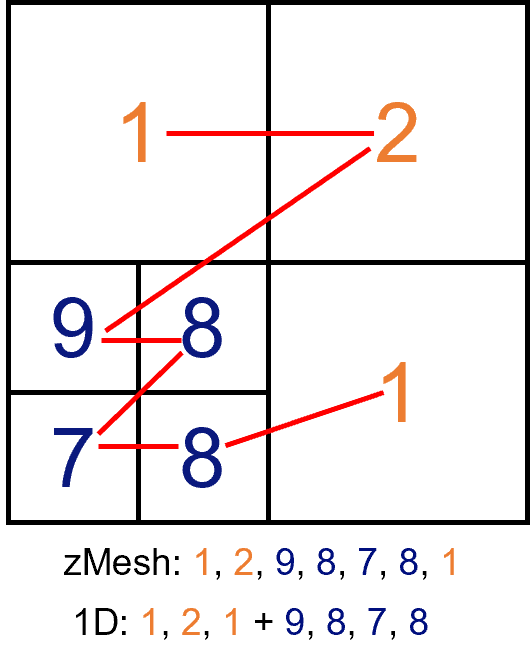} 
        \vspace{-5mm}
         \caption[t]{Tree-structured AMR data}
         \label{fig:ourdata}
     \end{subfigure}
     \begin{subfigure}[t]{0.45\linewidth}
         \centering
         \includegraphics[width=\linewidth]{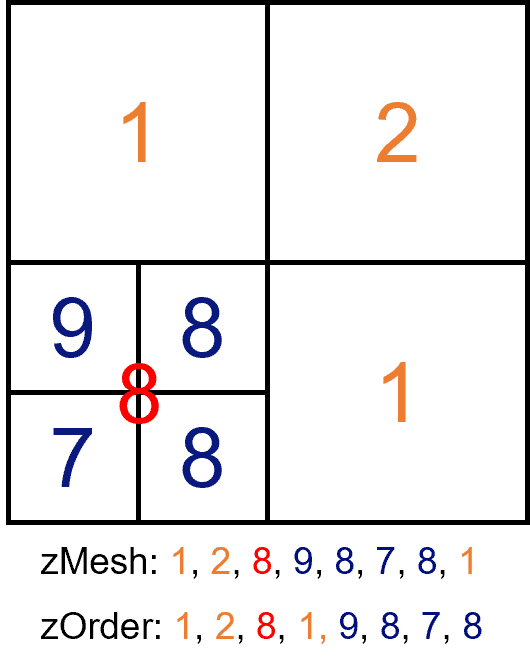}
        \vspace{-5mm}
         \caption{Block-structured AMR data}
         \label{fig:zdata}
     \end{subfigure}
        \vspace{-3mm}
        \caption[t]{An example of how the 1D baseline, zMesh, and original z-order reorder a simple 2D AMR data without and with redundancy. Orange: coarse level , blue: fine level, red: redundant data. 
        }
        \label{fig:zmesh}
\end{figure}

Thus, zMesh cannot improve the smoothness if there is no data redundancy in the tree-structured AMR datasets (i.e., our tested datasets). 
A simple example is used to illustrate this in Figure \ref{fig:zdata}, where the finer-level data has higher values because a grid will be refined only if its value is larger than a certain threshold. For block-based AMR, when a grid needs to be refined because of its high value, the value will still remain in the level, resulting in a redundant value saved (i.e., the red 8). 
If one uses the original z-ordering to traverse the data level-by-level (shown in Figure \ref{fig:zdata}), the reordered data will have three significant value changes (i.e., from 2 to 8, from 8 to 1, and from 1 to 9).

To solve this issue, zMesh traverses the two AMR levels together based on the layout of the 2D array. The reordered data are ``1-2-8-9-8-7-8-1'', which only has two significant value changes (i.e., from 2 to 8 and from 8 to 1). Thus, zMesh can improve the smoothness/compressibility for block-structured AMR data. 
However, as shown in Figure \ref{fig:ourdata}, for tree-structured AMR data (without saving a redundant ``8''), compared to the 1D baseline that compresses each level separately, zMesh introduces two significant data changes (i.e., from 2 to 9 and from 8 to 1) as it traverses between two AMR levels. This explains why zMesh is slightly worse than the 1D baseline on our tested AMR datasets.

When considering a 3D baseline, we found that it works slightly better than an adaptive compression approach.
First, from the high level, if the finest level of an AMR dataset has a very high density, it means that this dataset is not much different from a non-AMR dataset with uniform resolution. 
Thus, there is no need to use \textsc{TAC}. 
Instead, we can directly use the 3D baseline that up-samples coarse-level data and compresses the merged uniform data.   
This is because the main disadvantage of the 3D baseline is the redundant data generated by the up-sampling process; however, when the finest level is very dense, the coarse levels do not have much data to up-sample, thus the overhead of redundant upsampled data is almost negligible. On the other hand, compression on the uniform-resolution data (the 3D baseline) can better leverage the spatial information than the level-wise compression (\textsc{TAC}).

\begin{figure}[h]
    \centering 
    \includegraphics[width=\columnwidth]{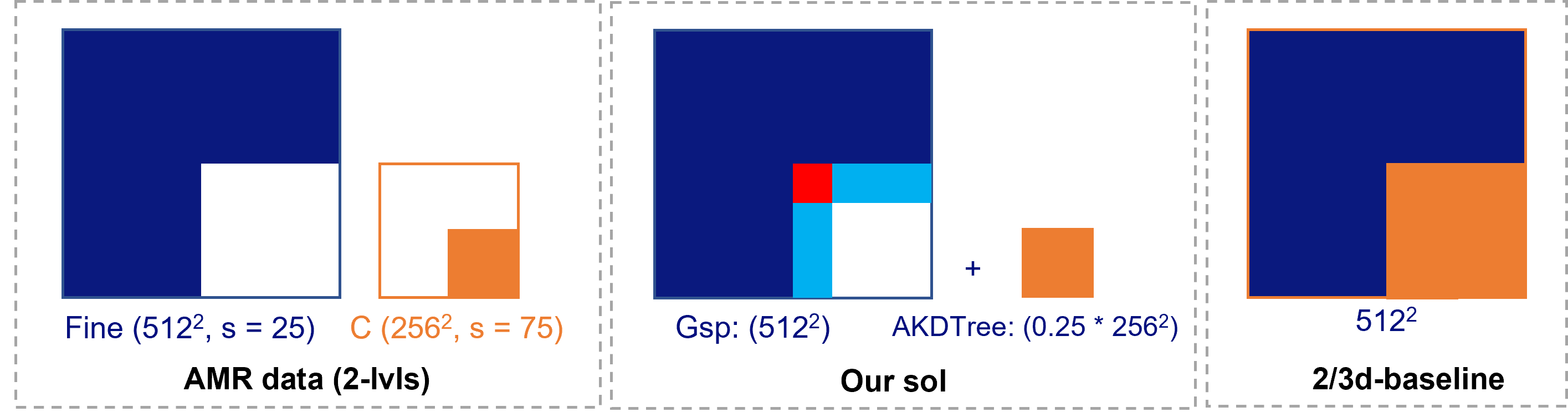}
    \vspace{-2mm}
    \caption{Comparison between the 3D baseline and \textsc{TAC} on an example AMR dataset with the dense finest level.}
    \label{fig:bse}
\end{figure} 

For an example of a two-level 2D AMR dataset as shown in Figure~\ref{fig:bse}. Its finest and coarse levels have the grids of $512^2$ and $256^2$, respectively.
If the density of the finest level is larger than 60\% (e.g., 75\% in the 2D example), \textsc{TAC} applies the GSP strategy to the finest level.
In that way, the data points to compress in the finest level and the coarse level are $512^2$ and $0.25 \cdot 256^2$, respectively. 
However, by simply using the 3D baseline, after up-sampling and merge, there are totally $512^2$ data points to compress. 
Therefore, instead of padding values to the finest level (i.e. GSP), we can simply fill in the up-sampled coarse levels to save the extra space of compressing the coarse levels separately and increase the smoothness/compressibility of the dataset.

Overall, we propose to adaptively use the 3D baseline and \textsc{TAC} based on the density of the finest level of an AMR dataset as follows:
(1) check the finest level's density; (2) use the 3D baseline to compress the data if the density meets the threshold $T_2$ we set, and (3) use \textsc{TAC} (OpST, AKDTree, and GSP) if the density does not meet the threshold.  

\subsection{Evaluation on Post-analysis Quality with Adaptive Error Bound}
\label{subsec:post-analysis}
We now evaluate \textsc{TAC} with the two cosmology-specific post-analysis metrics (i.e., metrics 5 and 6: power spectrum and halo finder) to demonstrate the benefit of the adaptive error bound method.
When factoring level-wise compression, \textsc{TAC} can apply different error bounds to different AMR levels based on (1) the post-analysis metrics, (2) the up-sampling rates of coarse levels, and (3) the rate-distortion trade-off between different AMR levels.
We choose the dataset run1-Z2 for evaluation because \textsc{TAC} is slightly worse than the 3D baseline on this dataset. 

\begin{figure}[h]
    \centering 
    \includegraphics[width=0.95\columnwidth]{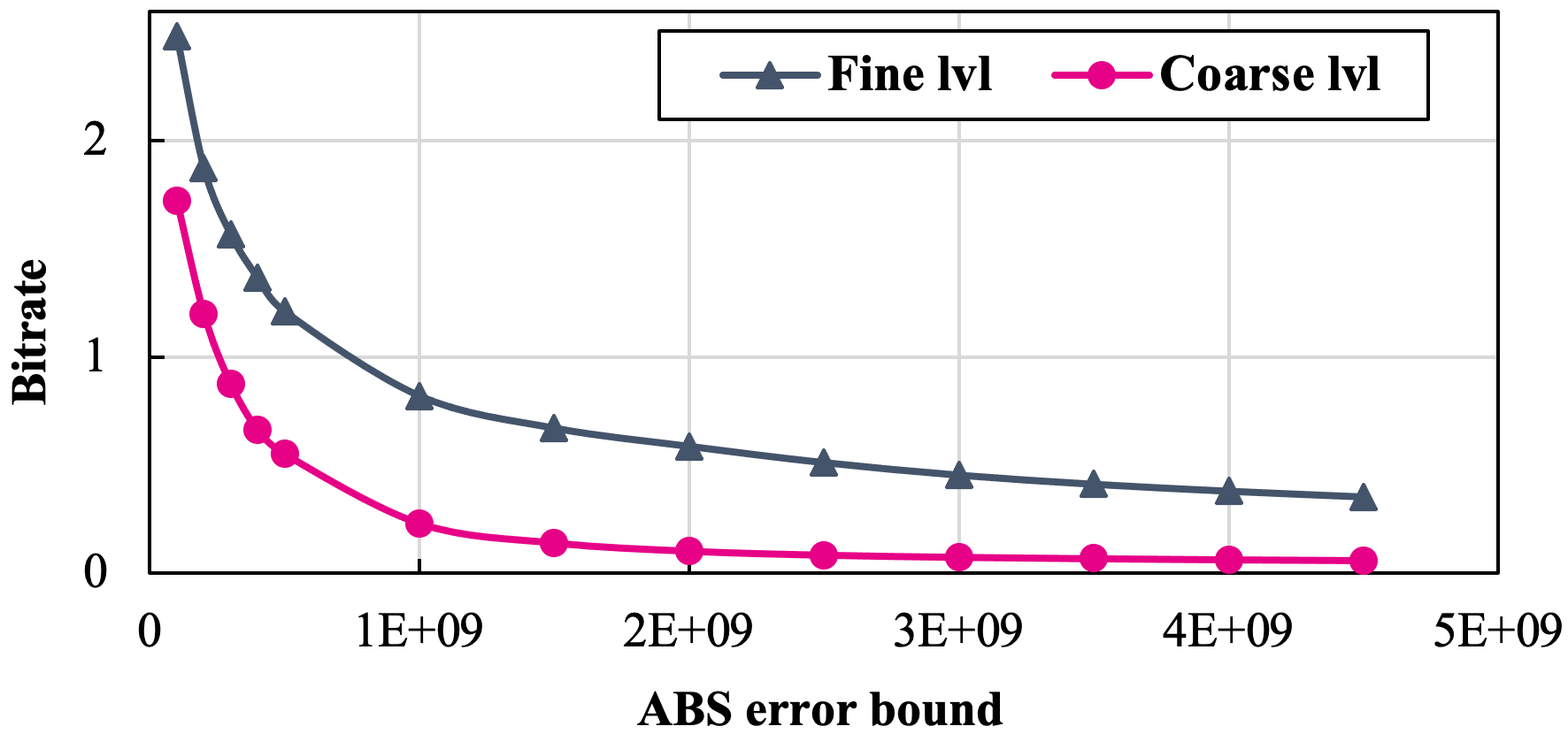}
    \vspace{-4mm}
    \caption{Bit-rates with different error bounds using SZ lossy compression for fine and coarse levels on Run1\_Z2 dataset.}
    \vspace{-6mm}
    \label{fig:db}
\end{figure}

Figure~\ref{fig:db} shows the motivation of performing rate-distortion trade-off between different AMR levels.
As the error bounds for the fine and coarse levels increase, their bit rates will converge to a similar value. This means that when the error bound is relatively large, the reduction in data size will be insignificant compared to the compression error increment (i.e., the slopes of both curves are very small). Therefore, we can say that when the error is large, it is not worth trading data quality for size reduction.


\paragraph{Power Spectrum}


\begin{figure}[h]
    \centering 
    \includegraphics[width=0.98\columnwidth]{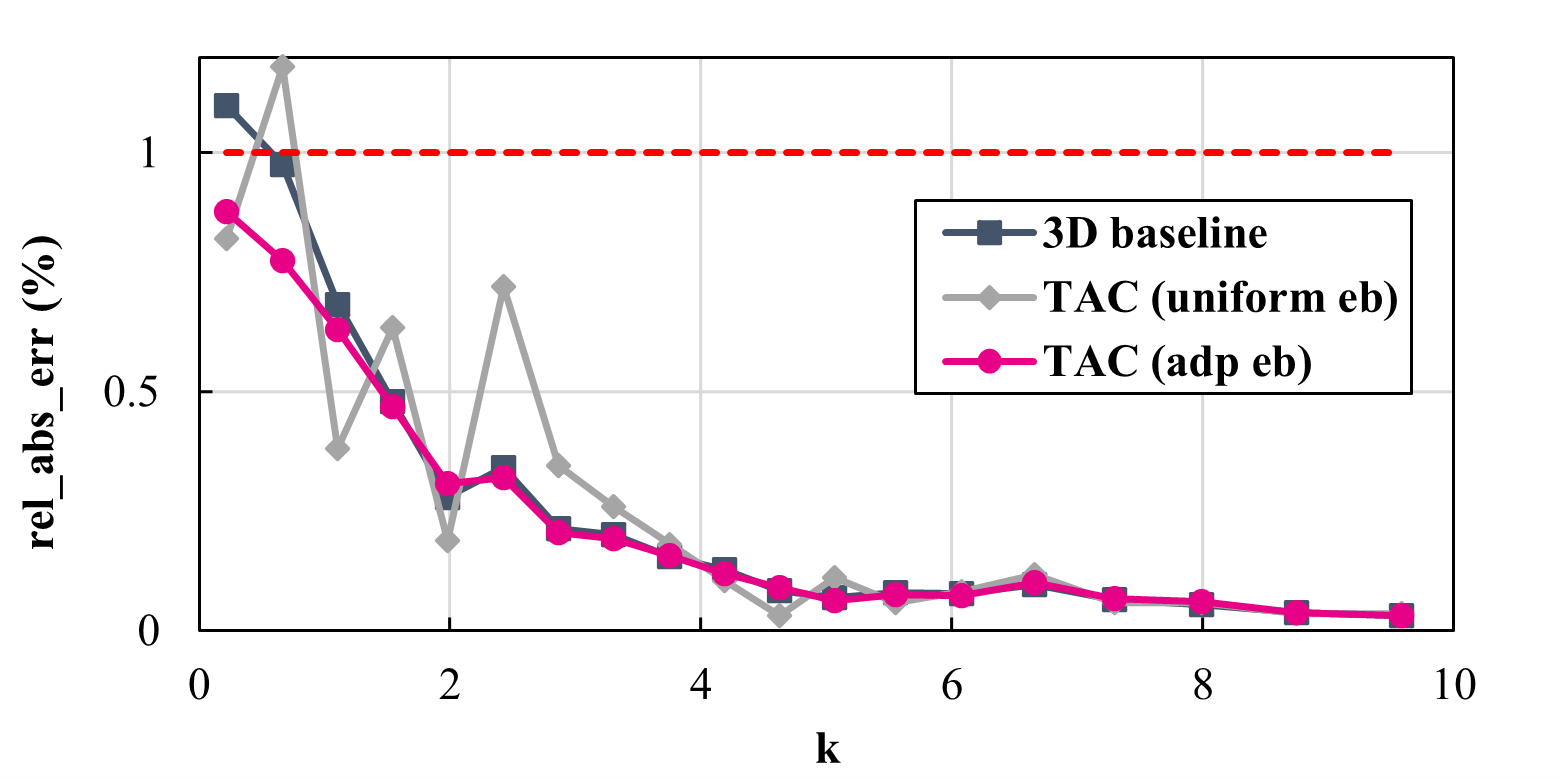}
    \vspace{-2mm}
    \caption{Power spectrum error (in relative) of the 3D baseline and \textsc{TAC} (the same error bound for all AMR levels) and \textsc{TAC} (different error bounds for different AMR levels) on baryon density field on run1-Z2. The red dashed line is the 1\% limit of acceptable power spectrum error.}
    \label{fig:ps}
\end{figure} 

Figure~\ref{fig:ps} shows that, under the (almost) same compression ratio, \textsc{TAC} (with the uniform error bound) has a similar power-spectrum error compared to the 3D baseline.

\begin{table*}
\caption{Overall compression/decompression throughput (MB/s) of different approaches with different absolute error bounds.}
\vspace{-2mm}
\resizebox{\linewidth}{!}{%
\input{Table/overall-throughput.tex}
}
\label{tab:tp}
\end{table*}

Now, let us follow the three steps mentioned at the beginning of this section to adjust the error bound for each AMR level.
First, the post-analysis metric--power spectrum---needs to be run on the uniform-resolution data and focuses on the global quality of data. Thus, the ideal error-bound configuration/ratio for the fine and coarse levels on the uniform-resolution data would be 1:1.

As aforementioned, the coarse level of the AMR dataset needs to be up-sampled to uniform the resolution. As a result, the compression error of the coarse level will be up-sampled as well, resulting in more error to the post analysis. 
Thus, we then need to give the coarse level a smaller error bound based on the up-sample rate. Here the up-sample rate for Z2's coarse level is $2^3$, leading to an ideal error-bound ratio of the fine and coarse levels changed to 8:1.

Finally, this 8:1 ratio needs to be adjusted based on the rate-distortion trade-off as aforementioned.
As shown in Figure~\ref{fig:ps}, when using the error-bound ratio of 8:1 (e.g., 4E+9 for the fine level and 5E+8 for the coarse level), the error bound of the fine level is too large, resulting in an ineffective rate-distortion trade-off.
Thus, we can balance two levels by increasing the error bound for the coarse level (to gain compression ratio) and decreasing the error bound for the fine level (to add compression error), which can achieve an overall rate-distortion benefit. Based on our experiments, we adjust the error-bound ratio from 8:1 to 3:1 and can observe that \textsc{TAC} has a significant improvement in the power spectrum error and outperforms the 3D baseline.


\paragraph{Halo finer} 
We evaluate the mass change, and the number of cells change for the biggest halo identified using the 3D baseline, \textsc{TAC} (with uniform error bound), and \textsc{TAC} (with adaptive error bound), as shown in Table~\ref{tab:hf}.
We can see that \textsc{TAC} with adaptive error bound produces better halo-finer analysis quality than the 3D baseline.

\begin{table}[h]
\caption{Halo finder analysis with different methods.}
\vspace{-2mm}
\begin{tabular}{|c|c|c|c|}
\hline
              & \textbf{CR}    & \textbf{Rel Mass Diff} & \textbf{Cell Nums Diff} \\ \hline
3D baseline   & 198.5          & 6.66E-04               & 39.00                   \\ \hline
\textsc{TAC} (1:1) & 198.5          & 4.97E-04               & 28.00                   \\ \hline
\textsc{TAC} (2:1) & \textbf{198.6} & \textbf{4.49E-04}      & \textbf{25.00}          \\ \hline
\end{tabular}
\label{tab:hf}
\end{table}

Similar to the error-bound configuration analysis done for power spectrum, let us now adjust the error-bound ratio between the fine and coarse levels for halo finder.
The halo-finer analysis also requires a uniform-resolution data as input. However, different from the power-spectrum analysis, the halo-finder analysis focuses more on high-value points in the fine level, since only high-value data points qualify as halo candidates, as described in Section~\ref{metric}. Note that this does not mean we can directly discard the coarse-level data with small values as they still contribute to the average value of the dataset, which is also an important parameter for the halo finder~\cite{Davis1985}. 
Therefore, we set the ideal error-bound ratio to 1:2 (i.e., fine level v.s. coarse level) for the uniform-resolution data based on our massive experiments.
After that, considering the up-sampling rate of $2^3$, the error-bounded ratio is changed to  4:1. 
Finally, we adjust the ratio to 2:1 based on the rate-distortion trade-off. Overall, as we can see in Table~\ref{tab:hf}, \textsc{TAC} with adaptive error bound obtains the minimal differences of the mass and cell numbers. 

\subsection{Evaluation on Time Overhead}
\label{subsec:throughput}
We evaluate the overall throughput (including pre-processing, compression, and decompression) on all the datasets with different error bounds. 
As shown in Tab~\ref{tab:tp} compared to the 3D baseline, the throughput of \textsc{TAC} is up to $75\times$ higher than on the Run2 datasets and $2.4\times$ higher on the Run1 datasets. 
This is because the Run2 datasets have lower density than the Run1 datasets in the finest level, resulting in a higher overhead of redundant data for the 3D baseline, which is consistent with our discussion in Section \ref{sec:dis}.
Moreover, \textsc{TAC} is slightly worse than the 1D baseline on the Run1 datasets due to the pre-possessing overhead. 
While we note that on the T3 and T4 datasets, our throughput drops due to a relatively heavy launching time (for compressing multiple 4D arrays generated by OpST) compared to the overall time on the small-sized datasets. 
Note that we exclude zMesh during the evaluation as it is theoretically slower than the 1D baseline due to the extra z-ordering and provides worse rate-distortion according to our evaluation. 







%% file: Table/dataset.tex
\newcommand\alignmiddle[2]{
\makebox[3em][r]{$(#1$}\makebox[.8em]{$,\ $}\makebox[2em][l]{$#2)$}}

\begin{tabular}{|l|c|c|c|}
\hline
\textbf{Dataset} &
  \textbf{\# Levels} &
  \textbf{\begin{tabular}[c]{@{}c@{}}Grid Size of Each Level\\  (Fine to Coarse)\end{tabular}} &
  \textbf{\begin{tabular}[c]{@{}c@{}}Density of Each Level\\ (Fine to Coarse)\end{tabular}} \\ \hline
Run1\_Z10        & 2 & 512, 256             & 23\%, 77\%                  \\ \hline
Run1\_Z5        & 2 & 512, 256            & 58\%, 42\%                  \\ \hline
Run1\_Z3        & 2 & 512, 256            & 64\%, 36\%                  \\ \hline
Run1\_Z2        & 2 & 512, 256            & 63\%, 37\%                  \\ \hline
Run2\_T2 & 2 & 256, 128            & 0.2\%, 99.8\%               \\ \hline
Run2\_T3 & 3 & 512, 256, 128       & 0.02\%, 0.56\%, 99.42\%     \\ \hline
Run2\_T4 & 4 & 1024, 512, 256, 128 & 3E-5, 0.02\%, 2.2\%, 97.7\% \\ \hline
\end{tabular}

\label{tab:datasets1}

%% file: Table/overall-throughput.tex
\begin{tabular}{|c *{7}{|c|c|c} | }
\hline
\multirow{2}{*}{$EB_{abs}$} &
  \multicolumn{3}{c|}{\textbf{Run1\_Z2}} &
  \multicolumn{3}{c|}{\textbf{Run1\_Z3}} &
  \multicolumn{3}{c|}{\textbf{Run1\_Z5}} &
  \multicolumn{3}{c|}{\textbf{Run1\_Z10}} &
  \multicolumn{3}{c|}{\textbf{Run2\_T2}} &
  \multicolumn{3}{c|}{\textbf{Run2\_T3}} &
  \multicolumn{3}{c|}{\textbf{Run2\_T4}} \\ \cline{2-22} 
 &
  {1D} &
  {3D} &
  \textsc{TAC} &
  {1D} &
  {3D} &
  \textsc{TAC} &
  {1D} &
  {3D} &
  \textsc{TAC} &
  {1D} &
  {3D} &
  \textsc{TAC} &
  {1D} &
  {3D} &
  \textsc{TAC} &
  {1D} &
  {3D} &
  \textsc{TAC} &
  {1D} &
  {3D} &
  \textsc{TAC} \\ \hline
1E+08 &
  {169} &
  {94} &
  97 &
  {166} &
  {90} &
  94 &
  {161} &
  {76} &
  99 &
  {160} &
  {40} &
  95 &
  {152} &
  {17} &
  76 &
  {143} &
  {2.4} &
  60 &
  {125} &
  {0.4} &
  30 \\ \hline
1E+09 &
  {219} &
  {115} &
  121 &
  {213} &
  {120} &
  127 &
  {208} &
  {109} &
  123 &
  {208} &
  {63} &
  117 &
  {193} &
  {27} &
  91 &
  {184} &
  {3.9} &
  66 &
  {159} &
  {0.5} &
  32 \\ \hline
1E+10 &
  {259} &
  {125} &
  135 &
  {256} &
  {125} &
  136 &
  {253} &
  {117} &
  137 &
  {250} &
  {65} &
  135 &
  {242} &
  {30} &
  102 &
  {229} &
  {4.0} &
  72 &
  {197} &
  {0.5} &
  34 \\ \hline
\end{tabular}


%% file: tex/06_conclusion.tex
\section{Conclusion and Future Work}
\label{sec:conclusion}

In conclusion, this paper proposes an error-bounded lossy compression for 3D AMR data, called \textsc{TAC}. It leverages 3D compression for AMR data on a systemic level. We propose three pre-processing strategies that can adapt based on the density of each AMR level.
Our approach improves the compression ratio compared to the state-of-the-art approach by up to 3.3$\times$ under the same data quality loss.
With our level-wised compression approach, we are able to tune the error-bound ratio of fine and coarse levels to be 3:1 and 2:1 for better power-spectrum and halo-finder analyses, respectively, under the same compression ratio.

In future work, we will apply our hybrid compression approach to more AMR simulations. We will also address the issue of relatively low throughput on small AMR datasets.



%

%% file: tex/99_acknowledge.tex
\section*{Acknowledgments}

This work has been authored by employees of Triad National Security, LLC which operates Los Alamos National Laboratory under Contract No. 89233218CNA000001 with the U.S. Department of Energy/National Nuclear Security Administration.
This research was supported by the Exasky Exascale Computing Project (17-SC-20-SC), a collaborative effort of the U.S. Department of Energy Office of Science and the National Nuclear Security Administration. This research was supported by the U.S. National Science Foundation under Grants OAC-2042084 and OAC-2104024. We would like to thank Dr. Zarija Lukić from the NYX team at Lawrence Berkeley National Laboratory for granting us access to cosmology datasets.